# Why optics needs thickness

David. A. B. Miller

*Ginzton Laboratory, Spilker Building, Stanford University, 348 Via Pueblo Mall, Stanford, CA 94305-4088, USA. dabm@stanford.edu*

We show why and when optics needs thickness as well as width or area. Wave diffraction explains the fundamental need for area or diameter of a lens or aperture to achieve some resolution or number of pixels in microscopes and cameras. Now we show that, if we know what the optics is to do, even before design, we can also deduce minimum required thickness. This limit comes from diffraction together with a novel concept called "overlapping non-locality" $C$ that can be deduced rigorously just from the mathematical description of what the device is to do. $C$ expresses how much the input regions for different output regions overlap. This limit applies broadly to optics from cameras to metasurfaces, and to wave systems generally.

## Introduction

Modern micro- and nano-fabrication let us make complex optics well beyond historic lenses, mirrors, and prisms, giving optics that does what we want, not just what previous optics offered. The resulting complex designs can, however, require long calculations, and might still be difficult to fabricate. The complexity also makes it hard to see in advance what may be possible. So, we want simple limits to guide us. For some optical function, what minimum sizes might we need, for example? From diffraction, we do understand how the minimum width or area of the optics must grow in proportion to the number of resolvable spots or pixels. We have had no corresponding basic understanding or limit on how thick the optics must be, nor even why optics fundamentally might require thickness.

Here we show why optics and other wave systems may need thickness and derive quantitative limits. Note that, if the optics is to do what we want, it may need to be "non-local" – the output at some point may need to depend on the input at possibly many positions. Such nonlocality means we need to communicate "sideways" *within* the structure or system. If we only need one "channel" for such communication, just a single thin layer may be enough. However, if the input position ranges for one output point need to overlap with those for another output point, we have what we call "overlapping nonlocality" (ONL). Any optical system beyond a simple transmissive or reflective mask may need ONL. A key realization here is that this ONL leads to thickness in optics.

We introduce this concept of ONL and define it as the required number $C$ of such "sideways" communication channels. A basic result is that the ONL comes just from the mathematical specification of what the device is to do. We can calculate $C$, quite rigorously, even before starting design. Then, with some heuristics from diffraction, we can deduce minimum thicknesses or cross-sectional areas for the optics from $C$. This approach gives limits for many optical components, including imagers and metasurface structures for a variety of possible applications. More generally, it bounds sizes for complex wave systems of any kind, including radio-frequency and acoustic systems.

Two recent questions in metasurfaces motivate this work now. First, can we shrink the distance between a lens and the output plane in an imager – "squeezing space" (*1*), possibly with a "space plate" (*2–5*)? Second, what kinds of mathematical operations could we perform – for example, on an image – using some metasurface structure with some thickness (*6–8*)? Our approach gives meaningful answers to these questions and others. It gives limits even for operation at just one frequency, so is complementary to a space-plate bandwidth limit (*4*) based on whether there is enough material to support the device's function (*9*, *10*) and to related semi-empirical limits (*11*). Limits in optics and electromagnetics are of increasing recent interest (*12*), and we believe our approach adds a new set of concepts, results and directions in this field. These results may find applications in other areas with complex optics, such as mode converters (*13–15*), and optical networks (*16*) in neural (*17–19*) and other (*20–22*) processing and interconnects (*23*).

Our optical system, Fig. 1A, takes the light on an input surface, and routes it to an output surface. We add a dividing surface that mathematically cuts through both the input and output surfaces; as it does so, it defines a "transverse aperture". We can deduce a minimum area or thickness for this aperture by counting the number $C$ of independent channels that must pass through it. For a camera or imager we can evaluate $C$ relatively intuitively. We also introduce a rigorous mathematical approach based on singular value decomposition (SVD) that applies to optical and wave systems generally.

## Overlapping non-locality for imaging systems

An imager might have a lens surface as its input and an array of pixel sensors as its output. We take it to be nominally loss-less – other than for incidental losses, such as weak background absorption, minor surface roughness scattering or reflection losses, it routes essentially all the relevant input power to the outputs. We also presume reciprocal optics – if waves can flow in one direction, then their phase conjugates can flow in the reverse direction, with the same transmission factor.

An imager takes a set of $N$ overlapping orthogonal inputs and maps them, one by one, to its $N$ separate output pixels (see supplementary text S1 for extended discussion and proofs). We presume, as is typical for an imager, that the input power for each output pixel is distributed essentially uniformly over the input surface.

We now divide both input and output surfaces mathematically in half with surface $S$ in the $y$-$z$ plane. Now, an imager is very much larger than a wavelength. So, we presume we can construct new approximate basis sets for each of half of the input surface, assigning a number of basis functions in proportion to the area of each part – so, $N/2$ input basis functions for each half. We presume that, in combination, this new "divided" pair of basis sets is approximately still able to describe all the $N$ orthogonal input functions.

Now consider the mapping between the *right* half of the input surface and the *left* half of the output surface (Fig. 1D); we now deduce how many orthogonal channels that mapping requires. Though $N/2$ orthogonal basis functions are associated with the right half of the input surface, we expect that half of those will be associated with forming images on the right half of the output plane. So, only $C_{RL} = N/4$ channels are associated with transferring power from the right half of the input plane to pixels on the left half of the output plane. Similarly, a number $C_{LR} = N/4$ of "left to right" channels are needed for waves from the left input surface to the right output surface.

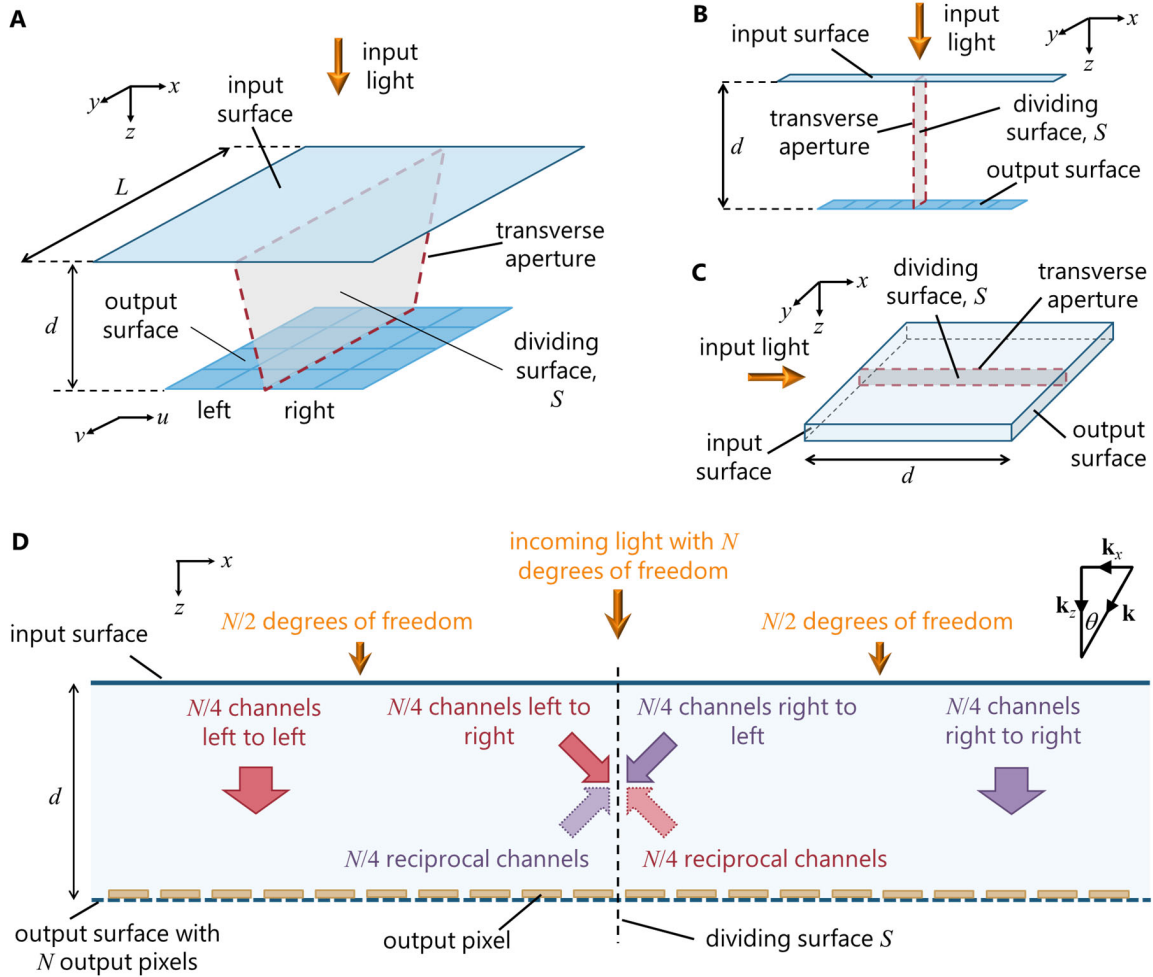

Fig. 1. Imaging systems and relevant surfaces and channels. (A) The input surface of an imaging system and the corresponding array of pixels on the output surface. The details of the optical system between these surfaces are not shown; we presume only that they are separated in $z$ by some distance $d$. A dividing surface $S$ that cuts through both input and output surfaces defines a "transverse aperture". (B) and (C) A one-dimensional imager, viewed either as a vertical slice, (B), that is thin in the $y$ direction and has "thickness" $d$, or, (C), as a thin slab in the $y$ direction and length $d$ in $x$, as in a photonic integrated circuit. (D) Required internal channels when dividing an imaging system with a large number $N$ of pixels and degrees of freedom into two equal parts. Coordinate systems for position and k-vectors are also shown.

Now, in deducing the total number $C$ of channels that must pass from right to left through the transverse aperture, we might think we could neglect any "left to right" channels because they are going in the other direction. However, by reciprocity, associated with those $C_{LR} = N/4$ "left to right" channels, there must also be an equal number of reciprocal or "backwards" versions of those channels from the output pixels on the right to the input surface on the left. So, altogether, we must physically allow for

$$C = C_{RL} + C_{LR} \quad (1)$$

channels crossing the dividing surface from right to left (or from left to right). In what follows, Eq. (1) applies quite generally. So, for our imager

$$C = N/4 + N/4 = N/2 \quad (2)$$

Note *C* here comes from the basic concept of how an imager must work and the number of pixels. It does not come from any specific design of the imager, nor even any specific sizes.

At this point, we can formally define ONL and *C*:

> The overlapping nonlocality (ONL) *C* associated with a dividing surface *S* passing through the input and output surfaces is the number *C* of orthogonal channels that must cross from inputs on one side of *S* to outputs on the other side of *S* to implement the desired optical function, summing over both directions ("left to right" and "right to left") of flow.

Next we deduce how large the transverse aperture must be to carry the *C* channels through it.

## Required area or thickness of the transverse aperture

First, we presume propagating electromagnetic waves carry these *C* channels. Presuming distances much larger than a wavelength, we neglect near-field electromagnetic terms (*14*). We presume simple "local" dielectrics – the polarization at some point depends just on the field at that point – so we neglect any nonlocality from plasmons or other compound excitations. So, we can use wave diffraction heuristics to predict size limits. For simplicity, we effectively consider just one electromagnetic polarization, but the same results would apply to each polarization.

We start by pretending the space between the input and output surfaces contains a uniform dielectric of refractive index $n_r$ with light of free-space wavelength $\lambda_o$. Diffraction heuristics (see supplementary text S2 for an extended discussion) tell us that in a narrow "slit" aperture as in Fig. 1B the maximum number of channels through the aperture corresponds to one for every $\lambda_o / 2n_r$ of distance in the *z* direction. If this space is not just a uniform dielectric, we can conjecture $\lambda_o / 2n_{max}$ per channel, where $n_{max}$ is the maximum refractive index in this space.

Finally, practically, we may be limited to using only some fraction $\alpha$ ($\leq 1$) of the full 180° range of angles inside the structure – equivalently, just a fraction $\alpha$ of the available k-space (i.e., of the component $k_z$ as in Fig. 1D) – reducing the available channels proportionately. For example, if the internal angle is restricted to a range 0 to $\theta$ as in Fig. 1D, then $\alpha = 1 - \cos\theta$. Hence, we conjecture in this one-dimensional (1-D) case that we need a thickness

$$d \geq \frac{C\lambda_o}{2\alpha n_{max}} \quad (3)$$

We can extend this heuristic argument to the area *A* of a two-dimensional (2-D) transverse aperture as in Fig. 1A, proposing

$$A \geq C \frac{1}{\alpha^2} \left( \frac{\lambda_o}{2n_{max}} \right)^2 \tag{4}$$

where now we regard $\alpha^2$ as the fraction of the 2-D $k_x, k_z$ k-space we are practically able to use in design. Equation (4) is equivalent to requiring an area of at least $(\lambda_o / 2\alpha n_{max})^2$ for each channel through the transverse aperture.

## Minimum thicknesses for imagers and related optical systems

We now apply Eqs. (3) and (4) to imagers. For a 1-D imager with $N_x$ pixels in a horizontal line in the $x$ direction as in Fig. 1B or C, from Eq. (2) we have $C = N_x / 2$, so from Eq. (3)

$$d \geq \frac{N_x \lambda}{4\alpha n_{max}} \tag{5}$$

For a 2-D imager as in Fig. 1A, with $N$ pixels (so $C = N / 2$ from Eq. (2)) and some characteristic width or diameter $L$, so with transverse aperture area $A \sim Ld$, then from Eq. (4),

$$d \geq \frac{N}{2L} \frac{1}{\alpha^2} \left( \frac{\lambda}{2n_{max}} \right)^2 \tag{6}$$

One subtle point for 2-D systems is that, to exploit the transverse aperture area effectively as in Eqs. (4) or (6), we may need to "interleave" degrees of freedom originally in $x$ into the $y$ dimension in the transverse aperture. This "dimensional interleaving" (DI) (see supplementary text S3) is possible in optics, and we can design "supercouplers" to achieve it (see supplementary text S4), including devising limits for these. Many approaches to optics, including free-space propagation, conventional imaging systems, simple dielectric stack structures, and 2-D photonic crystals do not, however, appear to support DI. In such cases, the thickness of these 2-D systems may end up as the 1-D limit, Eqs. (3) and (5). We compare with specific designs for imagers and "space plates" in supplementary text S5, showing these limits are both obeyed and approached in existing optimized designs.

An imager is one example of "space-variant" optics – it obviously looks different at different positions in the input or output. Several other such systems, such as Fourier transformers (*24*), mode sorters (*15*), and connection networks more generally (*16*) can be analyzed similarly; see supplementary text S6.

# Overlapping non-locality for general linear optical devices

An imager or mode sorter has a pixelated output, simplifying counting. Many optical devices, however, have no such pixelation, with continuous functions on input and output surfaces. The kernel – the linear operator relating the field at output points to that at input points – may be more local, unlike the imager's "global" kernel; a spatial differentiator, for example, relates an output region to a small number of adjacent input regions (see Fig. 2). Some devices, such as spatial differentiators, may not be unitary. The kernel may not be symmetric left to right, and it may not be obvious where to put the dividing surface. Fortunately, a singular value decomposition (SVD) (*14*) approach is both compatible with our arguments so far and these other cases.

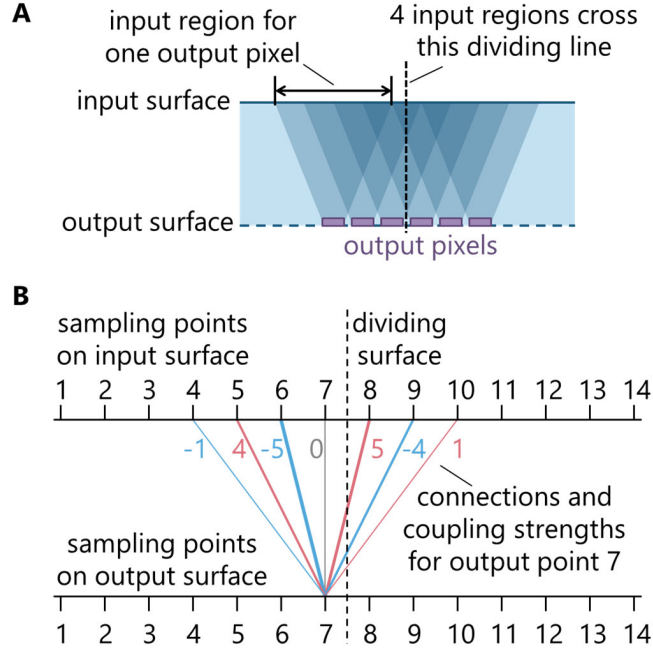

Fig. 2. Connections between input and output pixels. (A) A general example with an overlapping nonlocality (ONL) of $C = 4$. (The trapezoids between the input surface and the output pixels do not necessarily correspond to any specific optical beams or optical structures; they just show which pixel is connected to which of the overlapping input regions.) (B) Illustration of the coupling strengths between the input sampling points and output sampling point 7 for a central finite-difference approximation to a 5$^{th}$ order derivative. Similar coupling strengths apply for each output sampling point, just shifted sideways as appropriate as in (A). A dividing surface is shown between points 7 and 8 on both the input and output surfaces.

With coordinates $x$ and $y$ on the input face and $u$ and $v$ on the output face (Fig. 1), as in the formalism of Ref. (*8*), generally

$$\Phi(u,v) = \iint D(u,v;x,y) \Psi(x,y) \, dx \, dy \tag{7}$$

where $D(u,v;x,y)$ is the kernel or the "device" operator (*13*, *14*), relating the "output" function $\Phi(u,v)$ to the "input" function $\Psi(x,y)$.

Choosing a dividing surface at input and output positions $x_o$ and $u_o$, respectively, we have a "divided" operator $D_{RL}$ restricted to the "right" part of the input and the corresponding "left" part of the output,

$$D_{RL}(u,v;x,y) = \Theta(x-x_o)\Theta(u_o-u)D(u,v;x,y) \equiv \begin{cases} D(u,v;x,y) & u \leq u_o, x \geq x_o \\ 0 & u > u_o \\ 0 & x < x_o \end{cases} \qquad (8)$$

where $\Theta(z)$ is the Heaviside (or "step") function.

To find $C$, we start by finding the SVD of $D_{RL}(u,v;x,y)$. (Technically, we are establishing the necessary "mode converter basis sets" (*14*) to implement this "right to left" operator.) We then decide how many of the singular values (i.e., coupling strengths) have a large enough magnitude – i.e., above some small threshold – and use that as the number of required "right to left" channels, $C_{RL}$. If necessary, we set up a corresponding "left to right" operator

$$D_{LR}(u,v;x,y) = \Theta(x_o-x)\Theta(u-u_o)D(u,v;x,y) \qquad (9)$$

and similarly deduce a practical number of "left to right" channels $C_{LR}$; as above from Eq. (1), we add $C_{RL}$ and $C_{LR}$ to obtain $C$. For symmetric kernels, we may only need to calculate one of $C_{RL}$ or $C_{LR}$ and double it.

If, for some kernel, it is not obvious where to put the dividing surface, we could repeat the calculation for all reasonable choices of the dividing surface, and choose the largest result for $C$. As discussed in supplementary text S7, we should, however, keep the output "central" point $u_o$ "beneath" its corresponding input range.

## Constructing matrices for general linear optical devices

Because any such device operator *D* in a real physical system is necessarily bounded (i.e., finite output for finite input), *D* is necessarily a Hilbert-Schmidt operator, and hence also is compact (*14*, *25*); so, it can be represented to any precision by a sufficiently large matrix D.

Because our dividing surface is at specific positions in the input and output spaces, the matrix elements should give couplings between specific points in the input and output spaces. Effectively, these chosen points become "sampling points" for the functions in these spaces. With such "sampling" points, we can construct the matrix D. Matrices $D_{RL}$ and $D_{LR}$ are then just truncated versions of D; for example, for a 1-D problem, $D_{RL}$ and $D_{LR}$ are just the "upper-right" and "lower-left" quadrants of D. Then we can use standard matrix algebra to work out the SVD of $D_{RL}$ and, if necessary, of $D_{LR}$, and deduce *C* from Eq. (1). For pixelated optics, we could choose sampling points in the middle of each such pixel; essentially, we are then deducing limiting sizes for the optics so that it could give the right fields at least at these points.

For continuous functions and/or without pixelation, if we have no other constraints on choice of sampling points, then we just choose points with a separation "close enough" – intuitively, sufficient to represent even the smallest "bump" in the function. The criterion for "close enough" is then that the number of singular values of the resulting matrices, above some chosen small threshold of relative magnitude, has converged; so, further increasing the density of sampling points makes essentially no difference to the resulting *C*. Generally, experience in the SVD description of optics (*14*) shows this behavior quite consistently, with convergence guaranteed by the operator compactness and associated sum rules (*14*, *25*).

When the behavior of the optics depends only on the relative separation of input and output points, the optics is "space-invariant"; we are then convolving with a fixed kernel. Such convolution is common with metasurfaces (*8*). Then, *D* simplifies to

$$D(u,v;x,y) \to D(x-v, y-v) \tag{10}$$

Since the absolute position no longer matters, we simply choose one specific position for the calculations, e.g., for the output, such as $u=0$, $v=0$, and evaluate the matrices as required.

Much such metasurface discussion uses k-space (or Fourier) representations of functions and (space-invariant) kernels (e.g., (*1*, *5*, *6*)); "pixels" are not explicitly used. k-values must be smaller than $k = 2\pi n_r / \lambda_o$ for a propagating wave in the background material with refractive index $n_r$, or a smaller maximum value $k_{x\max} = 2\pi NA / \lambda_o$ if the input and output optics has a finite numerical aperture *NA*. In this case, we can use a "sampling theory" approach to get effective spatial sampling points. For details, see supplementary text S8. With *N* sampling points in one dimension, given the "bandwidth restriction" from the finite *NA*, these are spaced by

$$\delta l = \frac{\lambda_o}{2NA} \equiv \frac{L}{N} \tag{11}$$

where $L = N\delta l$ now becomes the finite nominal width of the surfaces for the purposes of this calculation.

## Example calculations of overlapping non-locality

### Pixelated systems

As a first explicit illustration of the full mathematical SVD approach, we consider here a device implementing a centered finite-difference 5[th] order linear derivative (*26*) in the *x* direction, in the spirit of Fig. 1B. Then 7 adjacent, equally-spaced sampling points would have weights proportional to -1, 4, -5, 0, 5, -4, and 1, as sketched in Fig. 2B for a dividing surface between points 7 and 8.

The connections between input points on the right of the dividing surface and output points on the left are expressed by the 3x3 matrix

$$\mathsf{D}_{RL} = \begin{bmatrix} 1 & 0 & 0 \\ -4 & 1 & 0 \\ 5 & -4 & 1 \end{bmatrix} \tag{12}$$

This matrix contains the connections from input points 8, 9, and 10 (corresponding to matrix columns) on the "right" to output points 5, 6, and 7 (corresponding to matrix rows) on the "left". All other connections across the dividing surface are zero, and so those possible additional rows and columns of the matrix are discarded; they would not anyway affect the SVD. (See supplementary text S9 and Fig. S7 for the full matrix $\mathsf{D}$ and matrices $\mathsf{D}_{RL}$ and $\mathsf{D}_{LR}$.)

Now we perform the SVD of $D_{RL}$. A standard numerical linear algebra calculation (here using the numpy Python library) gives the three singular values 7.568, 1.684, and 0.080, so $C_{RL} = 3$. If we similarly analyze the connections from left to right across the dividing surface, from input points 5, 6, and 7 to output points 8, 9, and 10, with this anti-symmetric kernel, the resulting matrix ends up being $D_{LR} = -D_{RL}^T$ (i.e., minus the transpose), as can be verified directly from Fig. S7, and has the same set of 3 singular values, giving $C_{LR} = 3$. So, for this 5$^{th}$ order finite difference derivative, we require $C = C_{RL} + C_{LR} = 6$ (as in Eq. (1)).

It might seem obvious that a $3 \times 3$ matrix with independent rows and columns has 3 singular values. However, we already see an important behavior we exploit later: not all the required channels are equally strong, and some may be negligible or nearly so. In fact, here, the 3$^{rd}$ channel in each direction is nearly $\times 100$ weaker than the first (0.080 compared to 7.568); this suggests that, if we only need a moderately good approximation for our derivative, we might get away with only 2 channels in each direction (so $C = 4$).

We can apply the same approach for other pixelated systems; finite impulse response filters and discrete wavelets, such as Daubechies wavelets, give additional examples (See supplementary text S10). For pixelated systems, in some simple cases, it is quite straightforward to understand ONL intuitively in optics that we can explicitly visualize; see supplementary text S11.

## Continuous systems

As an example of a continuous function as the kernel with such k-space limitations, we use

$$D(u;x) = \frac{(x-u)}{\beta} \exp\left(-\frac{(x-u)^2}{\beta^2 \Delta_t^2}\right) \qquad (13)$$

which is a real, 1-D version of the "$x$ times Gaussian" $\partial_x$ kernel of Ref. (7) that gives a "smoothed" differentiation. We allow for a "scale up" factor $\beta$ by which we can increase the distance scale of the kernel, with $\beta = 1$ corresponding directly to Ref. (7). As in Ref. (7), we take $NA = 0.15$, which by Eq. (11) leads to sampling points spaced by $\sim 3.33$ wavelengths, and $\Delta_t \simeq 8.325$ wavelengths. The resulting kernels for three different scales are shown in Fig. 3, together with the corresponding sets of relative singular values, including both right-to-left and left-to-right singular values in the same graph.

In each case, the full number of singular values equals the number of sampling points. After the first several singular values, however, the magnitudes of the remaining ones fall off very rapidly (we plot only the first 8 here). Note, too, that the set of strongly coupled singular values is essentially the same for all three scales of kernel. Once we have a large number of sampling points over the range where the kernel function is changing significantly, the relative size of the singular values converges. Note that increasing the numerical aperture also would not change the number of significant singular values, as long as the function is well enough sampled to start with. This illustrates that the ONL C is a property of the form of the function, not its scale, at least beyond some practical minimum scale. In all three cases shown, only the first 6 singular values have a relative size > 0.01. So, practically, we might choose $C = 6$ for this function.

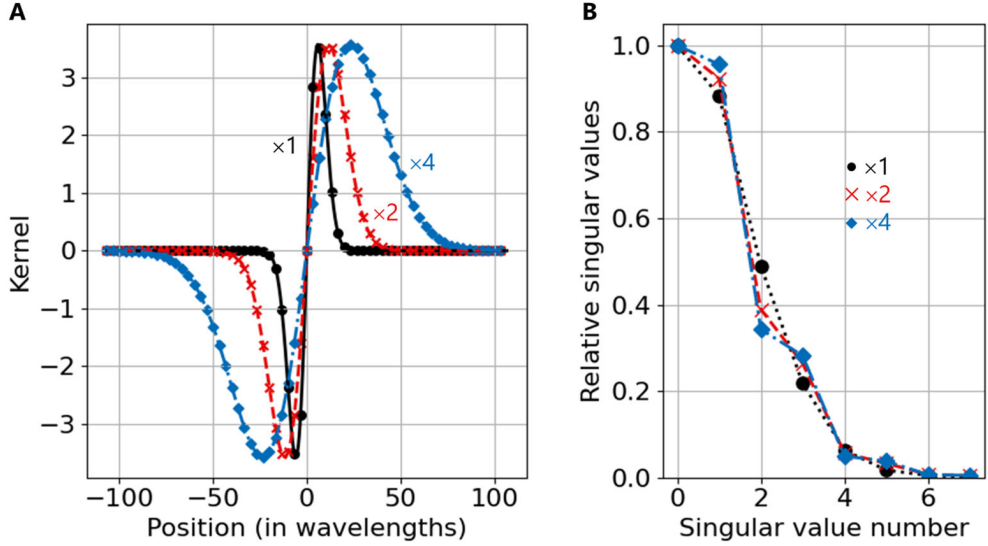

Fig. 3. An "x times Gaussian" kernel at three scales. (A) The kernel "×1" – the original scale ($\beta = 1$)) (circles and solid line), "×2" larger ($\beta = 2$) (crosses and dashed line), and "×4" larger ($\beta = 4$) (diamonds and dot-dashed line). The points correspond to effective sampling points for a numerical aperture $NA = 0.15$. (B) The corresponding relative magnitude of the singular values (including from left-to-right and right-to-left matrices) for the three scales of kernels (lines are just to guide the eye), using symbols and colors as in (A).

## Thicknesses for space-invariant kernels

These examples show many interesting discrete and continuous space-invariant kernels and operations that could be performed with values of $C$ from ~ 4 to ~ 8. Such numbers are likely still large enough that Eqs. (3) and (4) are usable at least as a first guide. (Some more sophisticated approaches using SVD are possible for thin structures and/or small $C$, without relying on the heuristics behind Eqs. (3) and (4); see supplementary text S12.) So, even without DI, such kernels might be implemented practically in structures that, for optical and near-infrared wavelengths, are only some small number of microns thick. A comparison with the "x times Gaussian" kernel design in (7) shows it also exceeds the minimum required thickness (supplementary text S13).

# Discussion

Our examples over a wide range of situations with waves, including pixelated, continuous, space-variant, and space-invariant systems show we now have a general method to predict minimum required thicknesses. We summarize the complete process and give some additional discussion in supplementary text S14 and S15, respectively. While we predict only fundamental minimum dimensions, we see above that systems with optimized designs already approach these limits within some small factor (e.g., ×3 or less). So, we now have a new and useful basic limit in wave systems based only on diffraction and straightforward mathematics.

**Acknowledgments:** The author is pleased to acknowledge stimulating conversations with Shanhui Fan.

**Funding:** DABM received funding for this work from the MURI program supported by AFOSR grant # FA9550-21-1-0312.

# Supplementary Text

## S1. Orthogonal channels in loss-less optics

Here we discuss in general the ideas of orthogonal channels and functions in loss-less optical systems and, in particular, for imagers with output pixels.

First, we should clarify what we mean by orthogonality. A simple case is that of two scalar fields $E_1(x,y)$ and $E_2(x,y)$ on some plane surface $P$ with coordinates $x$ and $y$. We could view these as electric fields in one specific polarization and frequency, for example. Then these two fields are orthogonal on this surface if and only if

$$\int_P E_1^*(x,y) E_2(x,y) \, dx \, dy = 0 \tag{1}$$

presuming, of course, that neither of these two fields is zero everywhere. (More sophisticated versions of orthogonality are also possible (*1*), including rigorous "power" or "energy" orthogonality with full vector electromagnetic fields, but the concept as in Eq. (1) will be sufficient for our discussions.) (As is common, for mathematical convenience we pretend the fields are complex numbers, here also with the presumption we are working with monochromatic fields, so of the form, $E \exp(i\omega t)$, with the implicit understanding that we can take the real part at the end of the calculation.)

We can note immediately that the fields on different output pixels are automatically orthogonal; the pixels do not overlap in space – pixels are separate areas on the output surface – so their overlap integral, as in Eq. (1) is automatically zero.

Quite generally, we can think of channels through linear optical systems as being pairs of functions – one function in the "source" or input space and the other member of the pair in the "receiving" or output space (*1*). Quite rigorously, there is a set of orthogonal functions in the source space that couples, one by one, to a corresponding set of orthogonal functions in the receiving space, in this pair-wise fashion; these sets of functions are known as the "communications modes" or "mode-converter basis sets". These pairs of functions can be thought of as the orthogonal "channels" in the optical system. These pairs of functions can be formally established by the mathematical process of singular value decomposition (SVD) of the coupling operator (Green's function) between the source and receiving spaces.

For loss-less systems, the SVD of the corresponding unitary operators is almost trivial (all the singular values are 1) and mostly reduces to a discussion of the dimensionality of the system. So, we will first construct more direct and intuitive arguments for such loss-less systems. One such argument for loss-less systems, which we can prove directly on physical grounds, is that the number of orthogonal channels or modes through a loss-less reciprocal optical system is conserved.

## Proof of conservation of "modal étendue", the number of modes in loss-less optics

It might even seem obvious that the number of orthogonal channels or modes through a loss-less reciprocal optical system is conserved, but I am not aware of an explicit published proof of this conservation law, so I give one here. This conservation could be viewed as a "modal" (*1*) version of conservation of étendue in loss-less optics (See also a related idea of "wave étendue" (*2*).) This result is helpful here as we set up the imager problem in the main text, though it may have other general uses in optics.

We consider a loss-less optical system, with *N* input modes, as shown in Fig. S1. In this proof, we use "single-mode" black bodies, a concept we have used in previous proofs (*1*, *3*). These are thermal black bodies that can only emit or absorb light in one single spatial mode, and over some identical narrow spectrum. (We could think of them as black bodies with a single-mode fiber as the sole input and output, and with a narrow-band spectral filter in that fiber.)

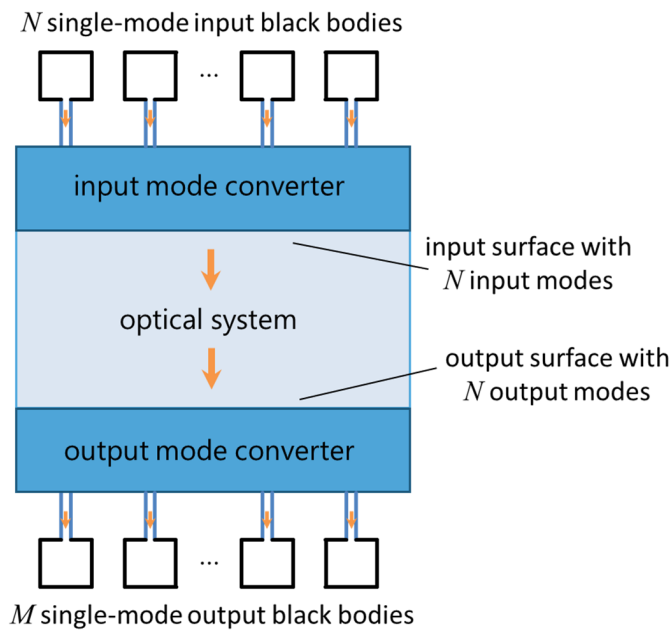

**Fig. S1 A thought-experiment apparatus connecting input single-mode black bodies through the loss-less optical system to output single-mode black bodies.**

Suppose, then, that the optical system of interest accepts all input power in *N* specific input modes. We then presume we construct an input mode converter that takes all the output power from *N* single-mode black bodies and routes that power, mode by mode, to these *N* input modes. (We know such an input mode converter can always be made; see (*4*) for a constructive proof.)

At the output, we take all the input power, which we now presume emerges in *M* output modes, and similarly convert it using an output mode converter to give the input powers to *M* single-mode black bodies at the output. We presume all these single-mode black bodies, at the input and the output, are at the same temperature.

Now, if $M < N$, then at least one such output black body must be receiving more power than it is emitting, which means that, starting from all black bodies being at the same temperature, we are now able to at least one of them up to a higher temperature. That violates the second law of thermodynamics. So, therefore, we cannot have $M < N$. We can next consider the system backwards, looking at the power emitted from the *M* "output" black bodies that is to be transmitted to the *N* input black bodies. By the same argument, we cannot have $N < M$.

Hence, we conclude that we must have $N = M$, which proves our assertion that the number of modes through a loss-less optical system is conserved.

This proof is stated so far as if for a reciprocal system. For a non-reciprocal system, we could argue that the backwards channels do not have to be the same channels as the forwards channels. However, if this is a closed system, a corresponding backwards channel has to exist for every forward channel in the system (*3*). In this case, the "input" and "output" mode converters would also have to be non-reciprocal, though again we know in principle how to do that (*4*). With this sophistication, the argument above also applies for non-reciprocal systems. So, this argument also applies for non-reciprocal systems provided they are closed in the sense of including these necessary backwards channels.

### Orthogonal channels and imagers

Imagers as discussed here are nominally loss-less systems. So, by the conservation theorem above, the number of orthogonal channels is conserved as we move through an imager. Such linear (nominally) loss-less optics can be represented by a unitary operator (at least within some overall loss factor). Unitary operators preserve orthogonality, and reciprocity means the optics is similarly unitary in the "backwards" direction. So, if the inputs are orthogonal, so are the corresponding outputs and *vice versa*. Since, by definition, pixels do not overlap, they correspond to orthogonal outputs, and hence the input fields associated with those output pixels are also necessarily orthogonal. Hence, as stated in the main text, an imager takes a set of $N$ orthogonal inputs and maps them, one by one, to its $N$ output pixels. (The orthogonal input functions associated with each output pixel are essentially uniform plane or spherical waves over the whole input surface.)

### Note on non-reciprocal systems

Note that, in the argument in the main text for counting the "sideways" channels in an imaging system, we presume for simplicity that the imaging system is reciprocal. If it is non-reciprocal, we might conclude that the "backwards" channels (e.g., from the output pixels on the "right" to the inputs on the "left") do not need to be included, at least not in the same form, and so might not have to pass back through the transverse aperture. If so, that would reduce $C$ by a factor of 2. As noted above, those backwards channels must exist somewhere, however, in any closed system. Conceivably, those non-reciprocal emitting channels could go out of the bottom of the imager, for example, which would allow us to reduce $C$ by a factor of 2. Whether in practice we could achieve such non-reciprocal separation of backwards channels without otherwise increasing the thickness of the imager system is an open question.

## S2. Extended discussion of the number of available channels through the transverse aperture

Here we give an extended and explicit discussion of the heuristic arguments leading to Eqs. (3) and (4) of the main text for the number of available channels through the transverse aperture.

In the 1-D case, as in Fig. 1B and C, we can think of the dividing surface in this case as a 1-D aperture of size $d$. We consider the possible orthogonal plane waves that "fit" in such an aperture, presuming for the moment that these waves are in a medium of refractive index $n_r$. With a free-space wavelength of $\lambda_o$, the magnitude of the k-vector inside this medium is $k_o = 2\pi n_r / \lambda$. The components of the k-vector in the *x* and *z* directions are $k_x$ and $k_z$ (see the inset in Fig. 1D). We allow both positive and negative values of $k_z$, corresponding to "downwards" and "upwards" waves, respectively, so in constructing a heuristic basis for waves propagating through this aperture, we allow only full periods to fit within the aperture width $d$. Hence, we have $k_z = 2m\pi / d$, where $m$ is an integer.

The largest possible value of $k_z$ is $k_o$, so the maximum magnitude of $m$ is $m_{max} = n_r d / \lambda$. We allow for the possibility that, for reasons of design, only some fraction $\alpha$ of the range 0 to $k_o$ for is practically available for $k_z$ (and, equivalently, for $m$ between 0 and $m_{max}$). Such a restriction could arise in some photonic crystal periodic structure because we can only effectively control some fraction of the Brillouin zone. Alternatively, there be some practical restriction on the maximum angle $\theta$ for the wavevector **k** inside the structure (see Fig. 1D inset). That in turn would give a minimum magnitude for $k_z$ of $k_{zmin} = k_o \cos(\theta)$, so a range of size $\Delta k \equiv k_o - k_{zmin} \equiv \alpha k_o$, so $\alpha = 1 - \cos\theta$, and similarly a range $\Delta m = \alpha m_{max}$ for the magnitude of $m$.

Allowing for this possible practical factor $\alpha (\leq 1)$, the total number of functions in this basis, and hence the number of orthogonal channels that can pass from one side of the surface to the other, is

$$m_{tot} = 2\Delta m = \alpha \frac{2n_r d}{\lambda} \quad (2)$$

where the factor of 2 is for the counting of positive and negative values of $k_z$ and, equivalently, $m$.

With no restrictions on the range of $k$ or internal angle, $\alpha = 1$, in which case $m_{tot}$ corresponds to the thickness $d$ of the structure measured in half-wavelengths inside the material – an intuitive result we might have guessed. It also corresponds to the maximum number of propagating modes in a slab waveguide of thickness $d$.

Now, in general, the material between the input and output surfaces will not just be some uniform dielectric. So now we make a conjecture that, in a structure with non-uniform refractive index, we can obtain a reasonable upper bound by replacing $n_r$ in Eq. (2) with the maximum refractive index inside the structure, $n_{max}$, giving

$$m_{totmax} \leq \alpha \frac{2n_{max} d}{\lambda} \quad (3)$$

for the maximum number of channels available in this 1-D imager problem across this surface from one side to the other. To be clear, we are not proving this conjecture (3); we merely assert it. (I am not aware of any exceptions to it, but that is hardly a proof.)

So, for $m_{totmax}$ to be large enough to support $C$ required channels, the vertical dimension $d$ of the slab in therefore must be

$$d \geq \frac{C\lambda}{2\alpha n_{max}}$$

as in Eq. (3) in the main text.

We can extend the kind of heuristic argument used for Eq. (3) in the main text with a similar argument for the $y$ direction. With a width $L$ in the $y$ direction, the "aperture" corresponding to the dividing surface would have an area $A \sim Ld$. Then we can similarly argue for an area limit (Eq. (4) of the main text) for the dividing surface of

$$A \geq C \frac{1}{\alpha^2}\left(\frac{\lambda}{2n_{max}}\right)^2$$

with $\alpha^2$ as the fraction of the 2-D $k_x, k_z$ k-space we are practically able to use in design.

## S3. Dimensional interleaving

There is a subtlety for optics with 2-D inputs and outputs: in practice, the thickness limit depends on whether we can use "dimensional interleaving" (DI). We can define DI as follows. Suppose the input field has $N_{ax}$ and $N_{ay}$ degrees of freedom (i.e., basis sets dimensionalities) in $x$ and $y$, respectively, that need to be communicated through the transverse aperture, giving a total number of degrees of freedom through the transverse aperture of

$$N_{atot} = N_{ax} \times N_{ay} \tag{4}$$

With corresponding coordinates $z$ and $y$ in the transverse aperture, and corresponding numbers of degrees of freedom $N_{bz}$ and $N_{by}$, by DI we mean that the optical system could reapportion these degrees of freedom between the physical dimensions, so that $N_{bz} \neq N_{ax}$ and $N_{by} \neq N_{ay}$, but still with the same $N_{atot}$, now given also by

$$N_{atot} = N_{bz} \times N_{by}. \tag{5}$$

Consider Fig. S2 as an illustration of these ideas. In Fig. S2A, we illustrate a case where the input field, possibly due to some restriction on the input numerical aperture NA (see supplementary text section S8 below), has some effective width $\delta l$ per degree of freedom in the $y$ direction that is relatively large; that is, $\delta l$ is presumed larger than the minimum possible width per degree of freedom inside the material of the structure, which would be $\lambda_o / 2\alpha n_{max}$ for some free-space wavelength $\lambda_o$, maximum refractive index $n_{max}$, and some practical restriction $\alpha \leq 1$ on the usable fraction of angles or k-space inside the device, as discussed in the main text.

For example, based on a "sampling theory" approach (see supplementary text section S8 below), we could have $\delta l = \lambda_o / 2NA$ for a numerical aperture $NA$, which is Eq. (11) of the main text. If that effective width $\delta l$ per degree of freedom in the $y$ direction is retained inside this 2-D device as in Fig. S2A, which is a case with no DI, then the number of degrees of freedom in $y$ is not changed, and $N_{by} = N_{ay}$. So, we need to use $N_{bz} = N_{atot} / N_{ay}$ degrees of freedom in the $z$ direction to get enough total degrees of freedom or channels through the transverse aperture. With a minimum thickness per degree of freedom of $\lambda_o / 2\alpha n_{max}$ in the $z$ direction, the area per channel through the transverse aperture is $\delta l \lambda_o / 2\alpha n_{max}$, and we now require a thickness, in $z$, of $d \geq N_{ax}\lambda_o / 2\alpha n_{max}$, which is the same result as a 1-D analysis with $C = N_{ax}$ degrees of freedom, as given explicitly in Eq. (3) in the main text.

Now, we could imagine that we could instead have some way of reapportioning the degrees of freedom between the two directions (i.e., DI), as in Fig. S2B, even if the input width in $y$ per degree of freedom, $\delta l$, is still the same. If so, we only need to allocate the minimum area $(\lambda_o / 2\alpha n_{max})^2$ per degree of freedom or channel through the transverse aperture. This would allow us to reduce the thickness $d$ of the structure. Note, though, that, inside the transverse aperture, we now have more degrees of freedom in the $y$ direction. Hence, we have somehow taken some of the degrees of freedom that were in the $x$ direction, and "interleaved" them into the $y$ direction; this is the hypothetical process of dimensional interleaving.

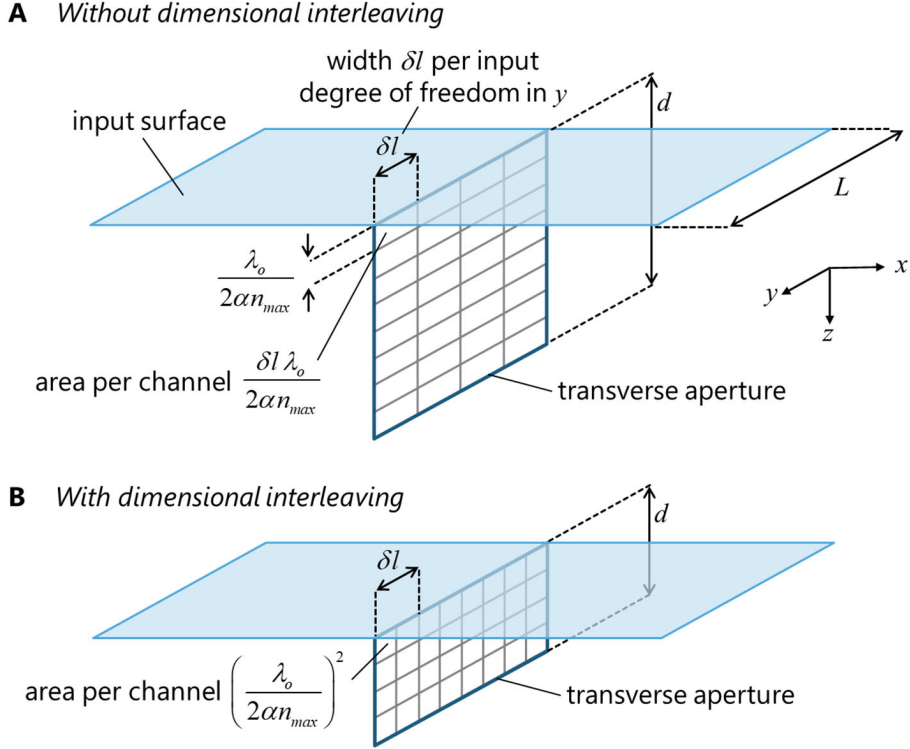

**Fig. S2. Illustration of degrees of freedom and effective areas per channel in the transverse aperture without and with dimensional interleaving. (A)** Without dimensional interleaving, with the $y$ width per degree of freedom, $\delta l$, set by the input optics and/or field, leading to the 1-D thickness limit even though this is a 2-D imager. **(B)** With dimensional interleaving (DI), though the input width in $y$ per degree of freedom, $\delta l$, may be the same, the optics somehow interleaves some of the $x$ degrees of freedom into the $y$ direction inside the structure, with, in this hypothetical example, now twice as many $y$ degrees of freedom used inside the structure. As a result of this DI, the thickness $d$ is reduced by a factor of 2 here. So, because of this hypothetical DI, the area per channel through the aperture is reduced, by a factor of 2 in this example, to the 2-D limit.

Incidentally, we are not necessarily saying that the individual channels explicitly "fit" one by one into these small rectangles in the transverse aperture as in Fig. S2, though, with wave-guided systems that is possible, e.g., with one waveguide per rectangle. But, if we look at one dimension – say, $y$ – the total number of degrees of freedom in the field within the device width $L$, for example for the input field, is $N_{ay} = L / \delta l$, which we could take as a definition of $\delta l$.

Most critically for a 2-D imager, we want to reapportion the $x$ and $y$ degrees of freedom between the $z$ and $y$ dimensions of the transverse aperture so we can make best use of its area. Now, changing degrees of freedom from $x$ to $z$ is simple – a 45° mirror is sufficient – but that gives no DI; we have not taken *x* (or *z*) degrees of freedom and "interleaved" them into *y* degrees of freedom.

Suppose we think of this optics in terms of the components of the incident *k* vector in the *y* direction, i.e., $k_y$ (as in supplementary text S8 below), with a set of $N_{ay}$ different $k_y$ values spaced by $\delta k = 2\pi / L$, as in a sampling theory approach. Then to proceed to the dimensionally interleaved situation as in Fig. S2B, we need to generate more $k_y$ values inside the structure – by a factor of 2 in this example. That means that the optical system inside the structure must be capable of generating such new $k_y$ components by some scattering mechanism. As we will see below, various forms of optics are *not* capable of doing this.

Dimensional interleaving is possible in optics. We could take a square (and hence 2-D) $M \times M$ array of single-mode fibers as the input, and have their outputs laid out side by side, so all in one "dimension" of with width $M^2$ – an $M^2 \times 1$ array. More generally, a device we could call a "supercoupler" could take *N* 2-D input modes and couple them into *N* side-by-side modes. (See supplementary text S4 below, where we also devise supercoupler limits.) Such coupling into just one "layer" or waveguide of thickness is the most extreme DI; this could give an imager that is only one such layer thick (though it would be impractically wide for any large number of pixels). We could, however, imagine other optical systems that did some degree of DI, with transverse aperture area still governed by Eq. (4) of the main text.

As mentioned in the main text, however, none of (i) simple propagation through free space (including also simple mirrors), (ii) conventional imaging systems, (iii) 2-D photonic crystal structures (at least if only using one band), or (iv) structures that are fully translationally invariant in *x* and *y* – so any structure made from multiple uniform layers, such as dielectric stacks (*5*, *6*) – appear to support DI. We discuss these in more detail below, but we can summarize the reasons here. For (i), free-space propagation, and (ii), conventional imaging, the core reason is that the propagation operator essentially remains separable in *x* and *y* (or, with a 45°mirror, in *z* and *y*). For (iii), 2-D photonic crystals, it is because, once coupled into the structure, any k-vector envelope function is actually a propagating eigenstate, and does not scatter into other such states. For (iv), translationally invariant structures, the transverse $k_x$ and $k_y$ components are conserved in a structure that varies only in *z*, so similarly they do not scatter into one another.

Without DI, we may need to revert to the "1-D imager" limit, Eq. (3) of the main text, rather than the 2-D limit of Eq. (4) of the main text, for the minimum thickness for a 2-D device with now $N_x$ as the number of pixels in one line in the imager's larger dimension (or in the longer "kernel" dimension in the space-invariant cases below).

## Specific approaches that do not support dimensional interleaving

Now let us examine in more detail these different classes of optics that do not appear to support DI.

*Separable kernels*

Presumeing $N_{ax}$ and $N_{ay}$ degrees of freedom, respectively, in the *x* and *y* directions in the input field, any optical input field can be described in the form

$$\Psi(x,y) = \sum_{p=1}^{N_{ax}} \sum_{q=1}^{N_{ay}} a_{pq} \psi_{xp}(x) \psi_{yq}(y) \tag{6}$$

where $\psi_{xp}(x)$ and $\psi_{yq}(y)$ are the corresponding basis sets in this input plane; Eq. (6) is simply a bilinear expansion on these basis sets. (Note here we write this field as if it is a scalar field for simplicity, but adding polarization merely corresponds to including polarization explicitly in the basis functions, and makes no difference otherwise in the expansion.)

Quite generally, we can write the field in the output plane as

$$\Phi(u,v) = \iint D(u,v;x,y) \Psi(x,y) \, dx\, dy \tag{7}$$

where $D(u,v;x,y)$ is the linear operator or "kernel" describing the relation between the output fields, with coordinates $u$ and $v$ (in the $x$ and $y$ directions in this argument), and the input fields, with coordinates $x$ and $y$.

For free-space propagation with some wavelength $\lambda_o \equiv 2\pi/k$, at least in the Fresnel approximation, the kernel is (*7*) (p. 120)

$$D(u,v;x,y) \propto \exp\left[i\frac{k}{2d}\left\{(u-x)^2 + (v-y)^2\right\}\right] \tag{8}$$

which we note is formally separable, being the product of two functions

$$U(u,x) = \exp\left[i\frac{k}{2d}(u-x)^2\right] \text{ and } V(v,y) = \exp\left[i\frac{k}{2d}(v-y)^2\right]$$

More generally, any such separable kernel can be written

$$D(u,v;x,y) \propto U(u,x) V(v,y) \tag{9}$$

Now suppose we expand one of these functions or operators, $V(v,y)$, bilinearly in orthogonal basis sets. Specifically, we can choose the basis sets $\alpha_q(y)$ and $\beta_q(v)$ that correspond to the singular value decomposition (SVD) of $V(v,y)$. Such an SVD is always possible for a finite linear operator; it yields these two unique sets of orthogonal functions, which are coupled, one by one, with coupling strengths given by the corresponding "singular values" $s_q$. Now, however, we know that, by choice, there are only $N_y$ degrees of freedom in the input field, so there are only $N_y$ orthogonal input functions $\alpha_q(y)$. So, necessarily, there are only $N_y$ orthogonal output functions $\beta_q(v)$ to which we can couple from the input. Formally, if we like, we can write the SVD of $V(v,y)$ as

$$V(v,y) = \sum_{q=1}^{N_y} s_q \beta_q(v) \alpha_q^*(y) \tag{10}$$

We could certainly imagine adding more orthogonal functions to the output space for functions of *v*, but none of them would be coupled to any input functions of *x*. We can argue similarly for $U(u,x)$.

So, if the kernel corresponding to this optical system is separable in *x* and *y* directions, then the dimensionality of the output spaces in *u* and *v* cannot be larger than the dimensionalities of the corresponding input spaces in *x* and *y*. So, with simple free-space propagation, we cannot perform "dimensional interleaving".

*Simple masks and thin optical elements*

A simple mask or thin optical element that is completely local (i.e., the output at a point is just dependent on the input at the same point) cannot perform DI; viewed on a "pixel" basis, for example, no information has been moved "sideways" in space at all. In the simplest view of imaging, in which we regard lenses as being thin elements in this sense, then the lens itself does not therefore perform any DI. So, in any optical system that can be viewed as a set of thin optical elements separated by free space propagation, at least in the Fresnel approximation, no dimensional interleaving is possible.

Incidentally, in this sense of a "thin" optical element, a simple plane mirror is also "thin". While it can certainly change the coordinate system for the optics – e.g., a 45° mirror could change input *x* and *y* coordinates to output $u = -z$ and $v = y$ coordinates – it still does not perform any DI.

*Two-dimensional photonic crystals and translationally invariant structures*

One approach for kernels that are space-invariant is to design a structure that is periodic in *x* and/or *y*, or more generally forms a 2-D photonic crystal (see, e.g., (*8*)). Such a structure repeats some unit cell to fill all the *x-y* space. This periodicity guarantees that the structure has the required space-invariance, at least for distance scales large compared to the unit cell size. The design of the system then reduces to the design of the unit cell.

From the Bloch theorem, such a system will have propagating wave solutions of the form

$$\Psi_{PC}(x,y,z) = Z(z) u_n(x,y) \exp(i(k_x x + k_y y)) \qquad (11)$$

where $Z(z)$ is some function of *z*, and $u_n(x,y)$ is a "unit cell" function that, for any given $k_x$ and $k_y$, is the same in every unit cell. When we use such a device, we expect that we will be coupling components of the input field with given $k_x$ and $k_y$ values into corresponding propagating modes in the photonic crystal with the same $k_x$ and $k_y$ values, and similarly for coupling out to the output field. A key point is that these waves as in Eq. (11) are eigenstates of the system. So, in propagation, they do not couple into one another. Whatever $k_x$ and $k_y$ values they start with, they retain. (There may in principle be some additional channels available in photonic crystals if we use unit cell functions from different bands (so a different index *n* in Eq. (11)), though such different modes correspond to different functions within a unit cell. Bloch functions in different bands also do not anyway scatter into one another in a perfect crystal.)

The structure shown below for a supercoupler (section S4 and Fig. S3B), which does perform DI, does have a periodic array of lenslets. However, the waveguide structure beneath it is *not* periodic in *x* and *y*, so this structure is not in any sense a photonic crystal in *x* and *y*.

Another approach that works for some space-invariant kernels is to use a structure that is completely translationally invariant – that is, it is exactly the same at each point in the *x*-*y* plane – so, a structure that varies only in *z*. One common class of such structures is "dielectric stacks" formed from successive uniform layers of two or more dielectrics (see, e.g., (*6*, *9*)). In such structures, the wave equation is separable, into an equation with *z* dependence and a simple wave-like equation in the *x* and *y* variables. Hence the $k_x$ and $k_y$ values of any incident plane waves are retained, just as for the photonic crystal case above, and different incident plane waves also do not couple into one another.

So, at least if we consider the basis sets for describing the input and output waves as being plane waves, as in a Fourier optics description, then the dimensionalities in the *x* and *y* directions are not changed by the interaction of these waves with both *x*-*y* translationally invariant structures and the 2-D photonic crystal (for incident waves that change little over a unit cell); no new k-states are created, and in particular no new waves with additional values of $k_y$ are created . This makes it difficult to see how such structures can usefully perform any DI.

## Conclusions on dimensional interleaving

The situation that is of most interest to us for reducing thickness in optics is to be able to take some of the degrees of freedom in *x* and dimensionally interleave them so we can usefully pack more degrees of freedom in *y* (if there is space) as we go through the transverse aperture; if we could do that, we could reduce the number of degrees of freedom in *z* (the "thickness" dimension) that otherwise have to accommodate the "*x*" degrees of freedom. As we have pointed out, this is indeed possible in optics using supercoupler approaches, for example.

But, we are concluding here that, at the very least as a first approximation, free-space propagation, imaging and Fourier transformations with lenses do not perform DI. We could argue that we have not proved this for "thick" systems, and that is quite correct. But we see that the underlying simple version of these processes certainly does not perform DI. As we have discussed, 2-D photonic crystal structures also would appear to have difficulty in performing DI, as would structures, such as multilayer dielectric stacks, that are fully translationally invariant in *x* and *y*.

## S4. Supercouplers and flat optical systems

We could imagine a useful "supercoupler" device that took in *N* overlapping modes in two dimensions and separated them to individual output waveguides or to lateral modes in a slab waveguide, all in a thin device, ideally just one "mode" thick (Fig. S3A). An array of grating couplers in the input area, each connected to one single-mode waveguide is one simple approach (Fig. S3B). There are many examples of such approaches (see, e.g., (*4*, *10–14*).

We discuss these here for two reasons. First, they are an explicit example of a device that performs "dimensional interleaving" (DI) (see the main text and supplementary text S3 above); though the inputs to such a device are two-dimensional – e.g., in *x* and *y* coordinates – the outputs can be essentially in one dimension such as *y* (Fig. S3B)– so the degrees of freedom in *x*-dimension in the input field are necessarily then "interleaved" somehow with the *y*-dimension input degrees of freedom into just the *y* dimension in the output. Second, with the approaches in this paper, it is straightforward to devise limits for supercouplers. This is relevant for the discussion of flat optical systems, so we start with this discussion of supercoupler limits.

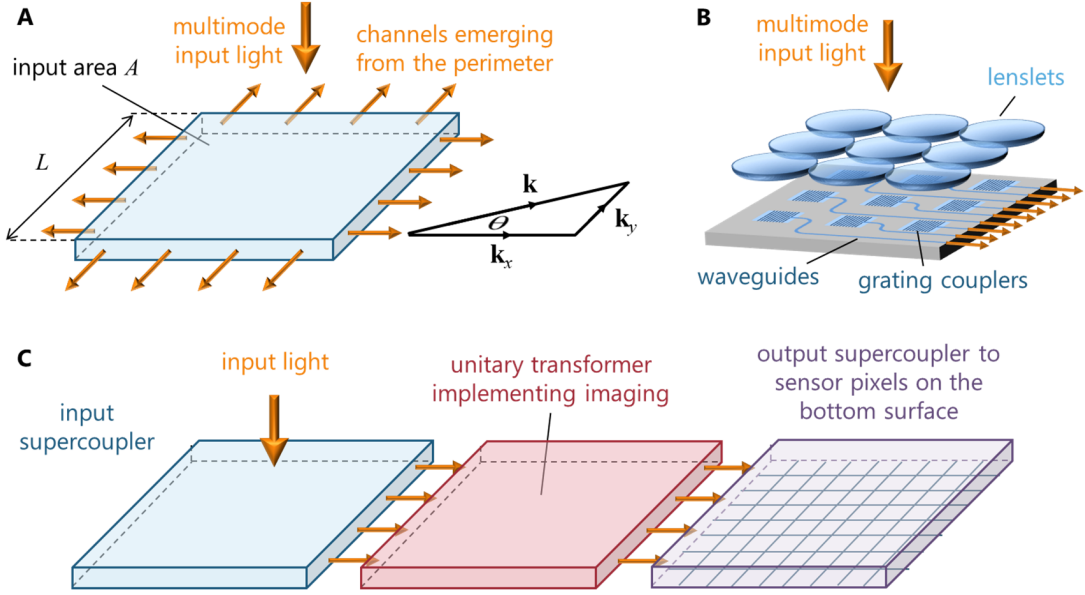

**Fig. S3. Super-coupler concepts and architectures. (A)** A "super-coupler" that couples from a number *N* of input modes on an area to a set of *N* output modes along a perimeter in a "thin" device, here illustrated as having a square input area of side *L*. **(B)** A hypothetical supercoupler design, with lenslets focusing into grating couplers that feed into waveguides emerging from one side. **(C)** A hypothetical "flat" imager.

## Supercoupler area limit

Consider supercouplers as sketched in Fig. S3A or B. The supercoupler is then some device that couples from the two-dimensional input field into modes or waveguides in a thin device, such as a slab waveguide or a photonic circuit of multiple in-plane waveguides. Fig. S3B sketches one explicit way in which this can be accomplished (*4*, *10*, *12*). Devices just with grating couplers and without the lenslets are also possible, with some reduced efficiency (*12*).

If the set of input modes of interest is spread equally in two dimensions, *x* and *y*, as in a nominally circular or square input area, this device has a minimum limit to its area because, by choice, we can only take outputs along the perimeter of the area. So, we are presuming explicitly that we are coupling to waveguides or modes in the plane.

As a practical matter, we may presume that either (i) we require single-mode waveguides emerging from a perimeter edge in Fig. S3A of B to have a center-to-center spacing that is $\gamma \lambda / 2n_{max}$ for some factor $\gamma > 1$ or (ii) there is either some maximum angular range $\pm \theta$ for waves emitted from a perimeter edge, so restricting the allowed "transverse" vector range by a factor $\sin \theta \equiv 1/\gamma$ (as illustrated for the $\mathbf{k}_y$ vector in Fig. S3A for modes emerging from the right surface). As a result, we require a perimeter length of

$$\Delta l = \frac{\gamma \lambda}{2 n_{max}} \qquad (12)$$

per mode or degree of freedom emerging from a perimeter edge, with such a practical factor $\gamma > 1$. (For example, in for silicon waveguides at $\lambda \sim 1.5$ microns using a refractive index $\sim 3.5$ and with a center-to-center spacing of 1 micron, we would have $\gamma \simeq 4.7$.) Then for $N$ input modes to be separated into those output waveguides, we need a perimeter of length at least $N\gamma\lambda / 2n_{max}$.

For a square input area of side $L$ that is to couple $N$ input modes to $N$ channels on the perimeter, as in Fig. S3A, then, with a perimeter of $4L$, we must have $4L \geq N\gamma\lambda / 2n_{max}$, i.e.,

$$L \geq N\gamma \frac{\lambda}{8n_{max}} \tag{13}$$

Hence, the area of the super-coupler must be

$$A \geq \left( N\gamma \frac{\lambda}{8n_{max}} \right)^2 \tag{14}$$

If the light is brought out of just one edge of the supercoupler (Fig. S3B), then the available perimeter is reduced by a factor of 4, giving

$$A \geq \left( N\gamma \frac{\lambda}{2n_{max}} \right)^2 \tag{15}$$

For a circular device, which has perimeter $\pi L$ for a diameter $L$, then (using all the perimeter) we would similarly have

$$A \geq \left( N\gamma \frac{\lambda}{2\pi n_{max}} \right)^2 \tag{16}$$

Note that these areas in Eqs. (14), (15) and (16) increase with $N^2$. So, as the number of such 2-D modes $N$ increases, the overall size of the input modes themselves must increase.

For example, though we could imagine focusing 100 orthogonal modes into a square area $\sim 100(\lambda/2n_{max})^2$ with some very good microscope objective, we could not couple all those modes out sideways into waveguides in a plane because there is not enough perimeter on that square to get the individual waveguides or modes out of the sides. According to Eq. (14), with our example we would need to have an area $A \geq (100\gamma\lambda/8n_{max})^2 = 625\gamma^2 (\lambda/2n_{max})^2$. Bringing the waveguides out of just one edge of the supercoupler would give, from Eq. (15), $A \geq (100\gamma\lambda/2n_{max})^2$. Even with $\gamma = 1$, this is an area 100 times larger than the area into which could focus 100 2-D modes, and with our practical example number of $\gamma \simeq 4.7$ for silicon waveguides on 1 micron centers, the supercoupler area would be $\sim 2000$ times larger.

The point here is not that supercouplers are uninteresting devices – they could be very interesting for many applications – but the requirements for getting the channels out of the edges does mean that their areas may have to be significant, and, for $N$ 2-D modes to be coupled, those areas scale with $N^2$.

## Supercouplers and "flat" imagers

Supercouplers also give us a way to imagine how we could make a completely "flat" imager or other kind of linear optical device. Suppose that we follow the in-plane output waveguides with an interferometer mesh (*4*, *10*, *12*) (Fig. S3C). Such meshes can perform any linear transform we like between input waveguides on the left and output waveguides on the right (*4*). For example, we could perform a unitary transformation with such a mesh, which allows us to construct any orthogonal linear combinations we like, therefore making an entire system that can separate $N$ such incident 2-D modes on any basis we like in the set of output waveguides.

Imaging is just a linear operation, transforming input beams of different angles or spherical curvatures into spots on output pixels. (The Supplement to Ref. (*15*) gives explicit examples of designs for such meshes, including for imaging operations.) Hence, in principle, we can use this approach to make a completely "flat" imager as in Fig. S3C, possibly even as thin as a single-mode slab waveguide if we can design suitable thin supercouplers.

As mentioned in the main text, however, the lateral dimensions of such a system rapidly become quite impractical as the number of pixels is scaled up. For example, if we imagined even a 1 MP imager with waveguides on 0.5 micron centers, the width of this device in the $y$ direction would have to be 0.5m just to accommodate the waveguides side by side. So, this flat imager approach would only be reasonable for relatively small numbers of pixels.

## S5. Comparison with existing imager and space plate designs

### An imager example

Consider, for example, a modern smartphone camera, presuming no dimensional interleaving (DI). With a 4:3 aspect ratio and effective resolution of 12 MP (megapixels) (*16*), the number of pixels in the longer, $x$ direction would be $N_x = 4000$. State-of-the-art lens designs for such cameras empirically do not appear to have "rays" at any angle larger than $\theta \sim 45°$ (*16*), giving a fraction of usable "$k_z$" space of $\alpha = 1 - \cos\theta \simeq 0.293$. Taking a typical maximum red wavelength in a camera as $\lambda = 700$ nm and $n_{max} = 1.5$, from Eq. (3) of the main text (as appropriate without dimensional interleaving), $d \geq 1.6$ mm.

Since smartphone cameras are practically restricted to ~ 5 mm total thickness, we can see that, with a lens design constraint of $\theta \sim 45°$, they are already within a factor of ~ 3 of this 1.6 mm limit. If, hypothetically, we increased the number of pixels by a factor of 2 in both directions, so a 48 MP camera with $N_x = 8000$, we would have $d \geq 3.2$ mm. So, though the geometric optics of lens design may still limit smartphone camera thickness (*16*), that thickness is approaching the limit given by Eq. (3) of the main text, and has certainly not passed it.

If we somehow presumed full DI, $\alpha = 1$, and with $L = 3.6$ mm for the shorter dimension on a 12 MP image sensor (*16*), then, from Eq. (4) of the main text we would obtain $d \geq 91$ μm. This extreme limit, which may be quite unrealistic practically, does, however, show that, even then we need significant thickness, possibly much larger than we would want to fabricate in a multiple layered metasurface structure.

## A space plate example

"Space plates" (*5*, *17–20*) would eliminate some or all of the length required for the "free-space" propagation between the lens and the image plane. To the extent that these work by emulating free-space propagation (but in some compressed distance), they would not give DI. So, as for 2-D imagers without DI, Eq. (3) of the main text would apply to the minimum distance between the input and sensor or image planes with such space plates. (Again, it would be possible in principle to make a space plate that did not work this way, and which internally did perform DI; in that case, we would use Eq. (4) of the main text to deduce minimum thickness.)

If a "space plate" design does not use DI, Eq. (3) of the main text also give limits for the minimum thickness $d$ to replace all the distance between the input and output surfaces. If we try to construct a space plate to shorten the 12 MP smartphone imager considered above, with, at best, $\alpha = 1$, our $N_x = 4000$ 1-D space plate thickness $d \geq 0.5 \, \text{mm}$.

Published work on space plates has mostly not considered the number of pixels explicitly, making comparisons of our limit with some of this published work difficult. However, Ref. (*20*) does explicitly simulate a full imaging operation with a space plate design, in their case with an output plane of 120 x 120 pixels. So the space plate design of Ref. (*20*) allows direct comparison with our limit, and shows good agreement, as we discuss next.

The structure in Ref. (*20*) is made from multiple uniform layers – a "dielectric stack" – so it is a "translationally invariant" structure. Hence, we do not expect it to support DI, as discussed in supplementary text S3. So, in one direction, there are $N_x = 120$ pixels, so $C = N_x / 2 = 60$.

The spacing from the input surface (a metalens, presumed thin) and the observation plane is $44.6 \lambda_o$, where $\lambda_o$ is the free-space wavelength; this space is entirely filled by the space plate of interest, which is composed of 50 layers of resonators. By our calculation that we need space for $C = 60$ channels, this device has enough thickness to support that at $\lambda_o / 2$ per channel even using the free-space wavelength (and the design in (*20*) is made from air-spaced structures). Equivalently, from Eq. (3) of the main text, using even $n_{max} = 1$, the required thickness would be $30 \lambda_o$, so close to, and smaller than, the $44.6 \lambda_o$ thickness of the design in (*20*).

That space plate design also explicitly has a structure with 50 channels if we consider each resonator layer as one channel, which is nearly enough explicit physical channels. The actual final image in this case in (*20*) has some aberrations and blurring, which could also be consistent with the system not quite being able to form all the pixels without overlap and/or with some blurring, and so it may have slightly less than 60 actual resolved pixels. Also, the image shown may not quite be filling the 120x120 pixel area.

Hence the space-plate structure in (*20*) is indeed consistent, both in thickness and in a direct counting of designed transverse channels, with the proposed limit as in Eq. (3) of the main text (so, without dimensional interleaving).

## S6. Other examples of space-variant optics and their thickness limits

In addition to the example of an imager in the main text, several other "space-variant" optical systems can be analyzed using the approaches in this work. Some can exploit the same analysis as used for the imager above. More generally, if that approach is not appropriate, we can use the singular value decomposition (SVD) approach, starting just from the mathematical function to be performed, as discussed in the main text.

## Systems with analysis similar to imagers

The arguments used for the imager in the main text would apply to essentially any efficient (i.e., low loss) optical system in which we have a pixelated output (or one that we can view that way), and in which the input "basis" functions corresponding to each output pixel are spread essentially uniformly (or nearly so) across the input aperture. Systems that are generally like this include optics that form Fourier transforms using lenses (see, e.g., Ref. (*21*)), and mode sorters that separate large sets of *N* (overlapping) modes to separate output spots (*22*). In those cases, for the 2-D limit (Eq. (4) of the main text), as for the imager, $C = N/2$, where *N* is the number of output pixels or spots. If there is no dimensional interleaving in the optical system, then we revert to the 1-D limit (Eq. (3) of the main text), with $C = N_x / 2$ where $N_x$ is the number of pixels in the larger dimension of the output pixel array.

Some approaches to mode sorting, such as multi-plane light converters (*22*) – use multiple successive Fourier transform optics. Each one of these stages would have to obey the limit for a Fourier transformer with *N* degrees of freedom. Such systems can, of course, be folded optically, using proportionately greater overall width while retaining the "thickness" of only one such Fourier transformer (*22*).

## General linear reciprocal systems

The analysis of imagers and similar systems uses a special feature of those systems that argues that only half of the input degrees of freedom on one side of the dividing surface have to communicate through the transverse aperture to the other side; the other half of the input degrees of freedom are presumed to be associated with outputs on the same side, and so we do not need channels to carry them through the transverse aperture. Hence, for a total of *N* degrees of freedom, for the ONL we have $C = N/2$.

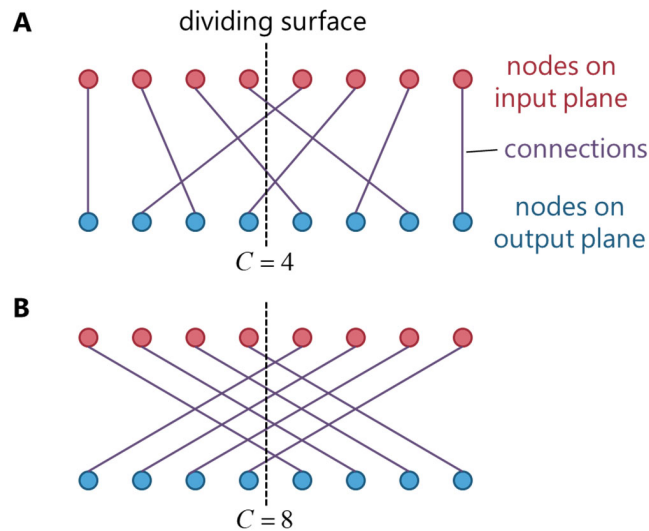

**Fig. S4. Example interconnection networks with 8 input and output nodes. (A)** (Perfect) shuffle network. **(B)** Part of the first stage of a butterfly network.

There are many topologies of interconnection networks. For some of these, we can see what value of *C* we require just by inspection. Figure S4 shows two simple examples as illustrations. Figure S4 A shows a (perfect) shuffle network. This has the same behavior as the imager in that half of the inputs on one side of the dividing surface connect on that side, and half connect on the other side. So, for *N* nodes on both sides, $C = N/4 + N/4 = N/2$. Figure S4 B shows another possible kind of interconnection network, in this case what is essentially the "non-local" part of the first stage of a "butterfly" network. In this case all the input nodes on one side connect to all the output nodes on the other side, so $C = N/2 + N/2 = N$.

Suppose, instead, that I want to propose a (reciprocal) optical structure in which I may be able to design any linear mapping between inputs on one surface and outputs on another. Given that I want to have that design freedom, what then is the minimum thickness the structure must have? To answer that question, we need to know how large *C* may have to be; if we know that, then we can calculate a thickness limit from Eqs. (3) or (4) of the main text.

In such a situation, there is a simple upper bound on *C*. First, should decide dimensionalities of the input and output optical spaces we want to use – so, the number of degrees of freedom or, equivalently, the dimensionality of the basis sets, $M_{in}$ and $M_{out}$ for the input and output spaces respectively. Now, we choose the smaller of those two numbers, calling it *M*; we will never have more than that number of independent channels between the input and the output. Next, we choose the "worst case", which is that the transverse aperture must be large enough to support *M* channels flowing through the aperture – for example, it may be that all of the inputs one the left have to connect to all of the outputs on the right (as in the first stage of a butterfly network in Fig. S4), and *vice versa* – not just half of them. This tells us the largest possible value of *C* we must allow for, which is $C = M$; this is the required upper bound on *C*. If we want to have the design freedom to construct an arbitrary linear optical device in this sense, then we require a minimum thickness as calculated using this *C*. Note that, among other things, this thickness limit would be required to allow us to be free to consider any *M*-dimensional interconnection network or, equivalently, any such directed acyclic graph (*23*).

## S7. Effect of displacing the output position

In the examples shown in the main text, we have taken the output "position" $u_o$ to be below the center of the input space of the kernel (whether space-variant, as in an imager, or space-invariant, as in convolution kernels), so directly below the middle of the input space for that output position. In part because those kernels have anyway had obvious symmetry (or anti-symmetry) about the center, this has seemed an obvious choice. This leaves open the question of what happens if we displace the output position, or even where to choose the output position if we have kernels that are not symmetric or anti-symmetric about some center. In the examples of causal finite impulse response kernels – the Daubechies wavelets in supplementary text section S10 – we chose the output just outside one end of the kernel.

The answer to this question of the effect of moving or choosing the output position $u_o$ is that, provided this position lies under or just beside its own input region, there is essentially no change in the resulting *C*. To understand why this is, we can look at Fig. S5, which illustrates a pixelated case as in Fig. 2A. Moving the output pixel sideways but "within" its own input range makes no difference to the number of regions crossed by the dividing line. However, if we move the output pixel by a distance *n* pixels (with $n = 3$ in this example) outside its own input region, so it is no longer below its own input region, then the dividing line crosses more regions, in this figure by a number equal to *n*. We can understand the physical reason for this behavior: we now need *n* more channels just to communicate the necessary output channels sideways by *n* pixels, "underneath" other input regions.

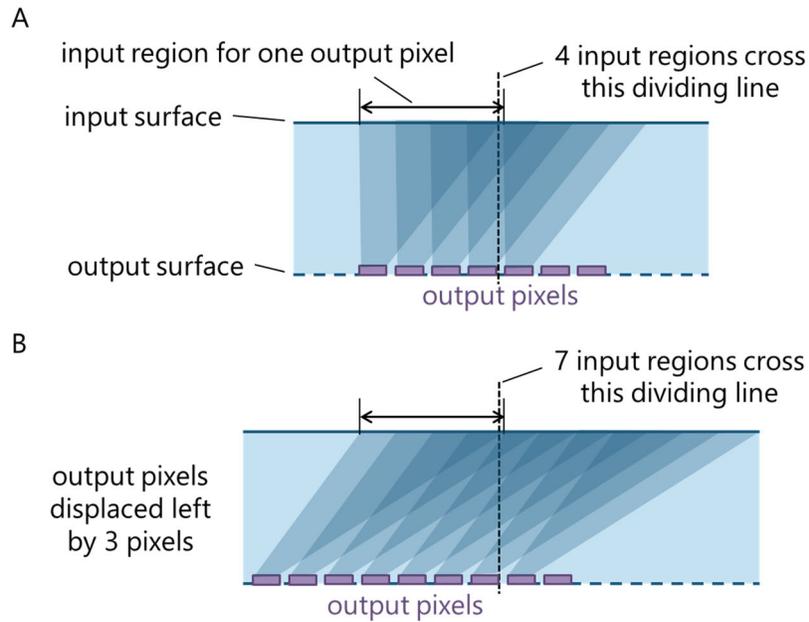

**Fig. S5. Illustration of the effects of displacing the output pixels. (A)** A situation as in Fig. 2A, but with the output pixel displaced to be below the left edge of the input range. This, however, does not change the number of regions crossed by the dividing line, which still crosses just 4 different input regions (it catches just the top right corner of the region from the left-most output pixel), so *C* is still 4 here. **(B)** With the output pixel moved by three pixels to the left, outside the input region, now the dividing line crosses 7 input regions.

In the case of continuous functions, which may have arbitrary numbers of input sampling points, possibly extending arbitrarily far out sideways, we cannot use the same simple "counting" argument as for fully pixelated case. However, we can still use the SVD approach to establish the number of channels we need, with strengths above some reasonable minimum strength.

Fig. S6 illustrates the results of calculations of the singular values for various displacements of the output position. In this illustration, we use the same "x times Gaussian" kernel as in Fig. 3, using the "×1" version (the original scale, i.e., $\beta = 1$ in Eq. (13) of the main text). We see that moving the output position from $u_o = 6\delta l$, which is just on the very right of the kernel as seen by eye, makes little difference to the set of singular values, and, in particular, the number of significant singular values, which remains at about 6 if we take $10^{-3}$ relative strength as a threshold. The kernel in this case does not have any particular overall width, decreasing but remaining finite as we move further from its center. Nonetheless, we see that, as we move out to a position of $u_o = 12\delta l$, the set of singular values has changed considerably, and there now are approximately 12 significant singular values. So, we are seeing similar behavior with the continuous kernel as for the pixelated kernel, with the number of required channels increasing as we move the output point outside the nominal width of the input region.

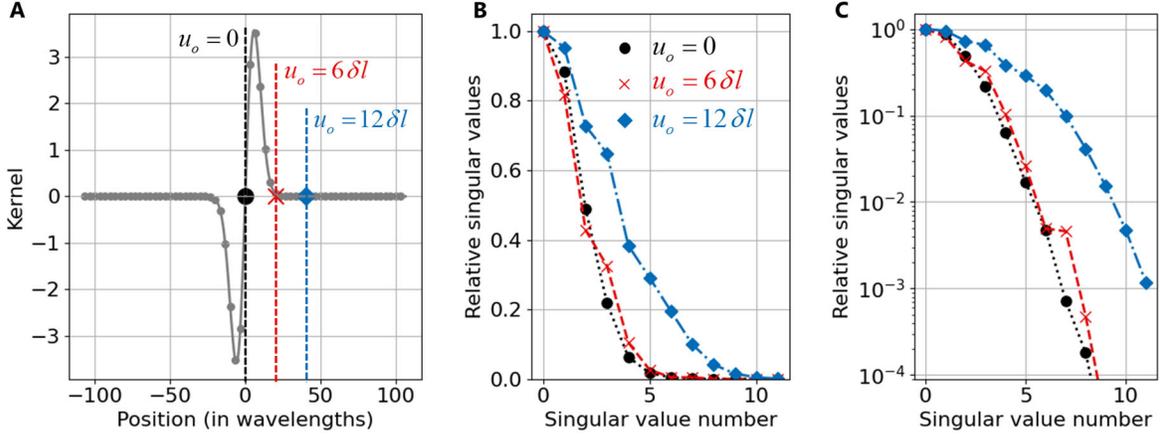

**Fig. S6. Illustration of the effects of displacing the output point for continuous kernels. (A)** The kernel (grey solid line) with the three displacement points indicated with a vertical dashed line and a corresponding point, $u_o = 0$ (circle), $u_o = 6\delta l$ (cross) and $u_o = 12\delta l$ (diamond), on the horizontal axis. (B) and (C) – relative magnitudes of singular values in linear and log scales, respectively, for the first 12 singular values. $u_o = 0$ (circles and solid line), $u_o = 6\delta l$ (crosses and dashed line) and $u_o = 12\delta l$ (diamonds and dot-dashed line). The illustrations here are otherwise for the same function and parameters as in Fig. 3 in the main text, using the same "x times Gaussian" kernel at the original "×1" ($\beta = 1$) scale) – with a numerical aperture $NA = 0.15$. Displacements are in units of $\delta l$, the separation between sampling points as calculated from Eq. (11) in the main text.

## S8. Sampling theory approach to spatial sampling points for device described in k-space

In considering (spatial) bandwidth bounds for optics described in k-space, obviously, the components $k_x$ and $k_y$ of the wavevector in the $x$ and $y$ directions must be less than the overall value $k = 2\pi n_r / \lambda_o$ for a propagating wave in a background material with refractive index $n_r$ and a free-space wavelength $\lambda_o$. It is also common to presume some specific numerical aperture (NA) for the optical system (see, e.g., (8)), which gives tighter limits on the set of allowed $k_x$ and $k_y$ values. NA is defined as $NA = n_r \sin\theta_{NA}$ where $\theta_{NA}$ is the maximum half-angle allowed for the optics in the input and/or output spaces. Then we must also have $\sin\theta_{NA} = k_{x\max}/k$, where $k_{x\max}$ is the maximum allowed magnitude of $k_x$. So

$$k_{x\max} = \frac{NA}{n_r}k = \frac{NA}{n_r}\frac{2\pi n_r}{\lambda_o} = \frac{2\pi}{\lambda_o}NA \qquad (17)$$

and similarly for the maximum magnitude $k_{y\max}$ of $k_y$.

For simplicity, we consider a 1-D problem, so with input and output coordinates $x$ and $u$ respectively, and we choose $u_o = x_o$ (a "vertical" dividing surface as in Fig. 1B). If we are interested in space-invariant kernels, then it does not matter what specific value we choose for $u_o$ (and hence also $x_o$). We take the physical system to have some finite width $L$ (which corresponds to a "repeat length" in our sampling theorem approach) that is much larger than the size of the "kernel" $D(u;x)$ in both $u$ and $x$. For a physical space of width $L$, running from 0 to $L$, we can choose $u_o = x_o = L/2$. With this chosen "bandwidth" of $\pm k_{x\max}$ for the k spaces, we need $N$ equally spaced sampling points in real space to specify this band-limited function, where $k_{x\max} L = \pi N$, i.e.,

$$N = \frac{k_{x\max} L}{\pi} = \frac{2L}{\lambda_o} NA \tag{18}$$

The k-space values are spaced by $\delta k = 2\pi / L$ and the real-space sampling points are spaced by

$$\delta l = \frac{L}{N} = \frac{\lambda_o}{2NA}$$

which is Eq. (11) of the main text, for both $x$ and $u$ values. (Technically, the points can viewed as running from to $-L/2$ to $(N-1)L/2N$ in real space, and from $-(N-1)k_{x\max}/N$ to $k_{x\max}$ in k-space, consistent with a common notation in discrete Fourier transforms to avoid double counting the "end points". With $N$ as an even number, these ranges include the value 0 as one of the points in both cases.) Hence, we can choose "sampling" points

$$u_q, x_q = q\delta l - \frac{L}{2} , \quad q = 0, 1, \ldots, N-1$$

With this discretization approach, we can analyze such k-space devices using the same matrix approach as for the pixellated or spatially sampled devices above.

Incidentally, we could also transform the real-space matrices $D$, $D_{RL}$ and $D_{LR}$ by discrete Fourier transforms into k-space representations $\tilde{D}$, $\tilde{D}_{RL}$ and $\tilde{D}_{LR}$, and work with those matrices. However, this makes no difference to the resulting singular values (we have also verified this numerically); it merely corresponds to a change in mathematical basis, with no change to the physical problem. Furthermore, the resulting matrices for $\tilde{D}_{RL}$ and $\tilde{D}_{LR}$ are larger, being essentially $N \times N$ in size, whereas the real-space matrix $D_{RL}$ is no more than one quadrant of such a matrix in size, and, as shown in supplementary text S9 below, is typically much smaller even than that.

## S9. Full matrix for the 5th order derivative example

The full matrix $D$ for the 5th order linear finite-difference derivative example in the main text is as shown in Fig. S7. We see this is a banded-diagonal matrix, with the width of the diagonal band in any one row or column corresponding to the number, seven, of coefficients in the corresponding kernel. The vertical dashed line corresponds to the position of the dividing surface on the input surface, and the horizontal dashed line to the position of the dividing surface on the output surface.

The matrices $D_{RL}$ and $D_{LR}$ correspond to the upper right and lower left quadrants of this divided matrix, respectively. For the purposes of calculation of the SVD of the matrices $D_{RL}$ and $D_{LR}$, there is no point in including rows or columns that consist entirely of zeros, so these matrices in practice then become the $3\times 3$ matrices as shown in Fig. S7.

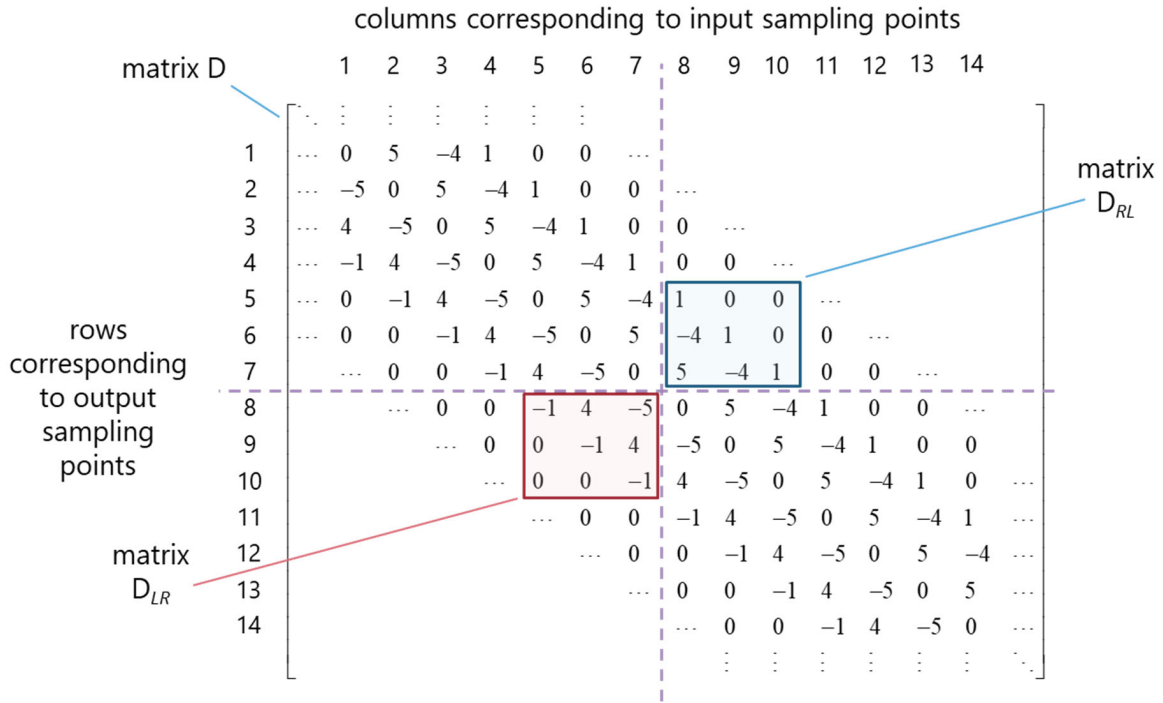

**Fig. S7. Full form of the matrix D for the 5<sup>th</sup> order linear derivative kernel.** Also shown are the two matrices $D_{RL}$ and $D_{LR}$ when the dividing surface is chosen between points 7 and 8 on both the input and output surfaces.

## S10. Overlapping non-locality for finite impulse response filters and discrete wavelets

As another "pixelated" or "discretized" example, we can consider finite impulse response (FIR) filters – a class of operations widely used in signal processing generally, including in image processing. Such filters work by adding up the sample amplitudes at a finite number of different relative points in the input function. Most simply, some finite number $m$ of sampling points is equally spaced in time or, in our example case, space in one dimension, $x$. We presume the input function (e.g., the input optical field) has values $f_0, f_1, f_2 \ldots$ at each of a set of equally spaced sampling points in this $x$ direction in the input plane. Then, for a device implementing such a FIR filter, the output at some corresponding pixel or sampling point on the output plane, indexed by some integer $q$, is

$$g_q = \sum_{p=0}^{m-1} a_p f_{q+p} \tag{19}$$

where $a_0, a_1, \ldots, a_{m-1}$ are the $m$ weights characterizing the filter response.

As two examples, we use the filter weights corresponding to two Daubechies wavelets, "db4" and "db12", as calculated using the PyWavelets Python package(*24*) at "level 1", which gives the shortest versions of these wavelets. These are plotted in Fig. S8A.

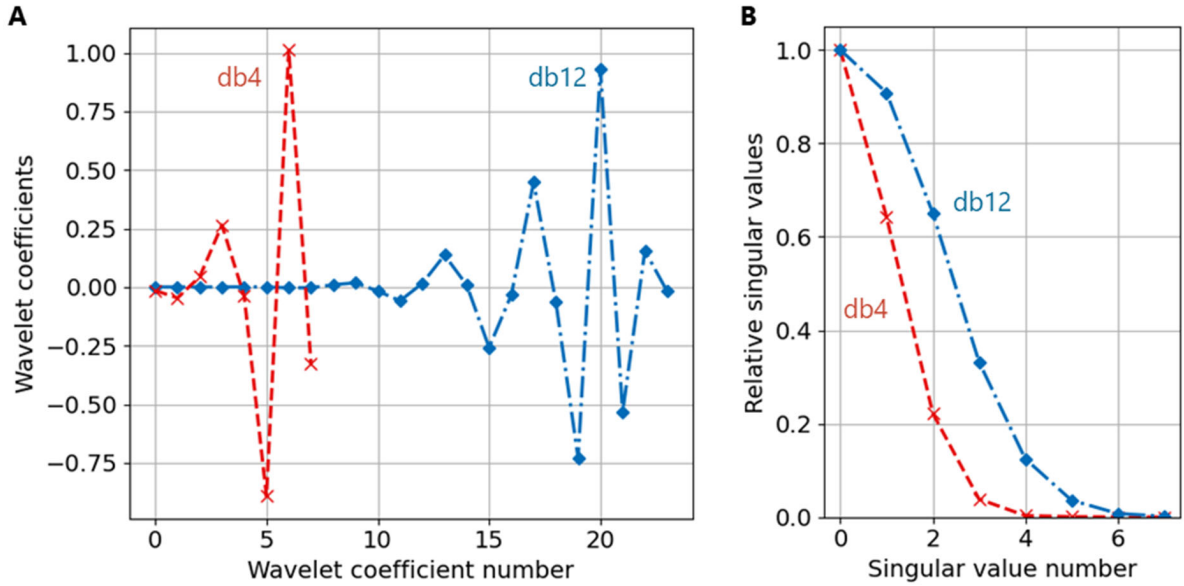

**Fig. S8. Coefficients and singular values for Daubechies wavelet kernels. (A)** Daubechies "db4" and "db12". (B). Relative singular values for the corresponding "left-right" matrix, with the output point just to the right of the corresponding input points. (Lines are guides for the eye only.)

There are $m = 8$ coefficients for "db4" and $m = 24$ for "db12". (Any "zero" coefficients at either end of the wavelet are trimmed off; they make no contributions.) In this case, we take the output point to be at the extreme right, just past the input wavelet, so we only need to evaluate the singular values for the $C_{LR}$ matrix, and that result will give $C$. The resulting singular values, relative to the strongest in each case, are plotted in Fig. S8B.

Numerically, as we would expect, there are 8 non-zero singular values for the "db4" wavelet, and 24 for the "db12" wavelet, numbers equal to the number of sampling points in each case, numbers that formally give us $C$ in each case; indeed, formally we would expect that $C = m$, the number of sampling points for any such FIR filter. However, in these examples, we find that, relative to the largest singular value in each case, for "db4" only 6 of these singular values are larger than $10^{-3}$, and for "db12", only 8. Hence, in practice, we may again be able to make a good approximation to this filter with correspondingly smaller values of $C$ in practice, such as 6 or 8 respectively.

## S11. Illustration of nonlocality concepts using discrete beam splitters

If we are considering optics in which both the inputs and outputs are effectively pixelated, we can propose conceptually simple ways of implementing devices with non-locality, including overlapping non-locality. These ideas may help understand some of the key concepts of the approach in this paper, so we illustrate these ideas here.

One conceptual approach uses networks of beam splitter blocks with added phase shifts (*4, 10*). We likely would not make an actual device in exactly this way, but this approach gives a tangible illustration of various of the concepts. Actual systems designed in this "building block" fashion can also be made practically using waveguide interferometer meshes (*4, 10*). Such meshes would themselves be quite large, at least as currently implemented, so they would not approach fundamental limits in thickness. However, the scalings of such interferometer mesh networks do correspond with the more fundamental ideas presented here.

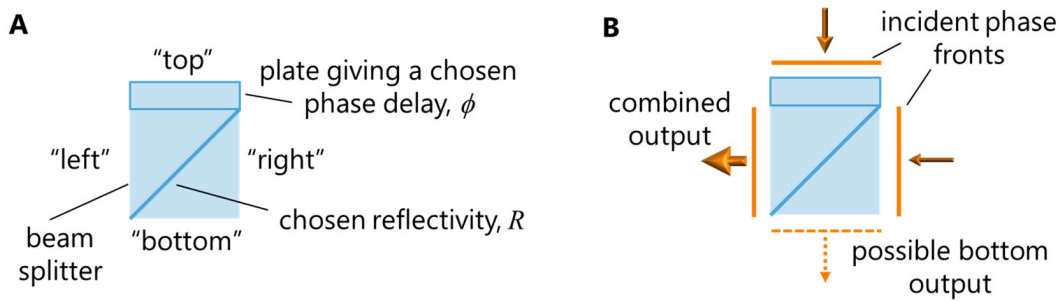

**Fig. S9. Illustration of a hypothetical "cube" beam splitter. (A)** A side view of a cube beam splitter, showing the reflecting surface on a 45° diagonal, with an additional plate on one face to give a phase delay. We presume the (power) reflectivity $R$ can be set to any desired value for such a beam splitter, and the relative phase delay between the "top" and "right" inputs can be set as needed by choosing the thickness of the plate.

First, consider just one "cube" beam splitter as in Fig. S9 A and B. Such a cube of glass has a reflecting surface inside it at 45° along a diagonal. Light incident on the top will be split to the left with some (power) fraction given by the (power) reflectivity $R$ of that surface. We also include a "plate" of some chosen thickness to give some additional phase delay $\phi$ to light incident from the top. We treat the top input face on the phase plate as an "input pixel" in these discussions.

In these discussions, we pretend, as in Fig. S9 B, that we can approximate light as plane phase fronts, each of approximately uniform intensity, in and out of the beam splitter on its various faces. At some frequency, for any given relative amplitude and phase of light incident on the top and on the right of the beam splitter, there is some choice of $R$ and $\phi$ that leads to all the light coming out of the left as a combined output power. (For light of other relative amplitudes and phases, in general we expect some light out of both the left and bottom faces.)

Now we imagine that we form a line of such beam splitter cubes, as in Fig. S10 A. Some incident light pattern of a given frequency shines on the set of beam splitters from the top. We pretend that we can approximate the light on each beam splitter top surface as a plane wavefront of a uniform intensity within the surface of a given beam splitter, though we presume different possible amplitudes and phases of these plane wavefronts on different beam splitters. We neglect any diffraction inside the beam splitters, so these plane wave fronts are retained throughout. (Again, if the reader is not enthusiastic about such approximations, we can substitute waveguide interferometer blocks fed from grating couplers on the top surface (*4*, *10*); the remaining arguments still hold.)

Now, there is some choice of the phase delays and reflectivities of the various beam splitters that results in all this input light being combined and routed out the left of this line of beam splitters. Then, if we put a 100% reflecting mirror at 45°at the left end, we can reflect all this combined output power into an output "pixel" on the bottom left. (This structure and operation is just a specific example of a "self-aligning beam coupler" as discussed in Ref. (*10*).)

Incidentally, with such schemes, it is easy to deduce just what reflectivities and phase delays we need if we imagine shining light backwards from the left through the line of beam splitters. That would in general result in light of specific amplitudes and phases coming out of the top surface;

that light is just the phase conjugate of the light that would be coupled "forwards" to come out of the left of the line of beam splitters.

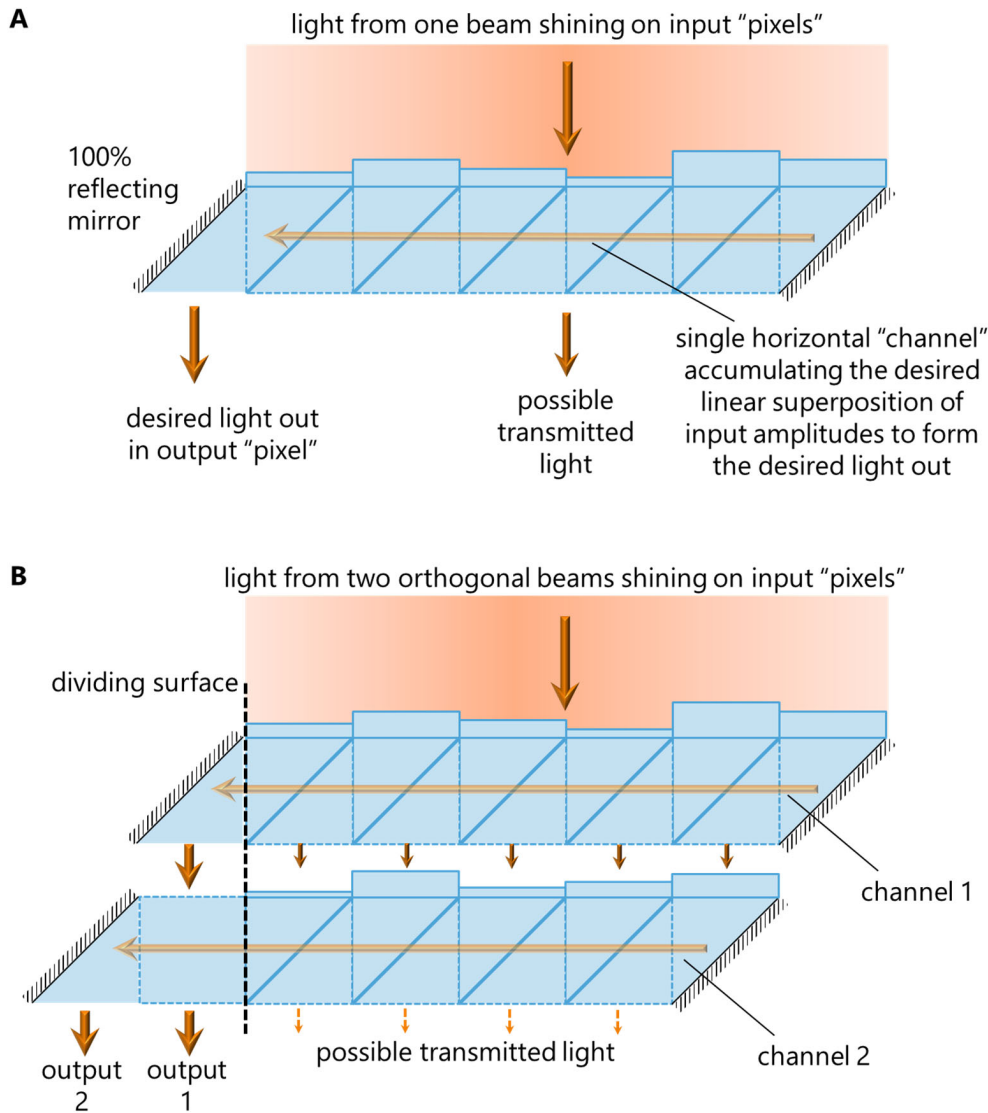

**Fig. S10. Illustration of lines of beam splitters for beam combination. (A)** A line of beam splitters that, with appropriate choices of beam splitter reflectivities and phase delays, can combine some specific input pattern of light shining on the top of the structure so it all comes out of the left of the line of beam splitters. We show this output being reflected so it comes out of one "pixel" on the bottom of the structure. **(B)** Two successive lines of beam splitters can separate two orthogonal input beams, routing them in two channels through the "dividing surface" to separate outputs.

Now, this line of beam splitters is behaving as a non-local device; light at multiple different input points is being combined to give one output. Note that we could extend this arbitrarily in our simple approximations just by increasing the number of beam splitters in this horizontal line; there is no limit to how many different such inputs can be combined to give one output. The thickness of this structure remains the same – essentially just one "channel" thick – no matter how much non-locality we require.

Now consider a structure as in Fig. S10 B that has two successive rows of beam splitters. In this case, the first row of beam splitters is the same as in Fig. S10 A. We presume these beam splitters are set so that all the light in a specific beam "1" comes out the left of the first row and is reflected to output 1. Suppose now we shine a second beam onto the top surface. This second beam we presume is "orthogonal" to the first. Mathematically, we can imagine we represent each beam by a vector of complex amplitudes – the amplitudes of the plane wave segments incident on each top beam splitter face. So we have a corresponding vectors $|\psi_1\rangle$ and $|\psi_2\rangle$ for these two beams. (This "$|.\rangle$" Dirac notation can be taken just to represent a column vector of numbers). By orthogonal we mean that the inner product of these two vectors is zero. Equivalently, we would have orthogonality directly in the fields, as in supplementary text S1 (Eq. (1)).

In this case, with orthogonal beams, none of the light in beam 2 can be coupled out of the left of the top row of beam splitters; doing so would violate the second law of thermodynamics – it would allow us to combine the power of two independent beams into one, and that would allow us to combine the power from two black bodies to heat up another one at the same temperature. So, all the power in the orthogonal second beam must pass, in some form, out of the bottom of the top row of beam splitters.

So, we can configure a second row of beam splitters to collect that light. With an appropriate set of phase delays and reflectivities in this second row, we can route all this light in the second beam out the left of this second row, and reflect it to come out of output 2 on the bottom. If we added some third input beam, orthogonal to the first two, then the power of that beam would be transmitted out of the bottom of the second row of beam splitters (where we could collect it in a third row of beam splitters, and so on). All this behavior is also discussed in Refs. (*4, 10*).

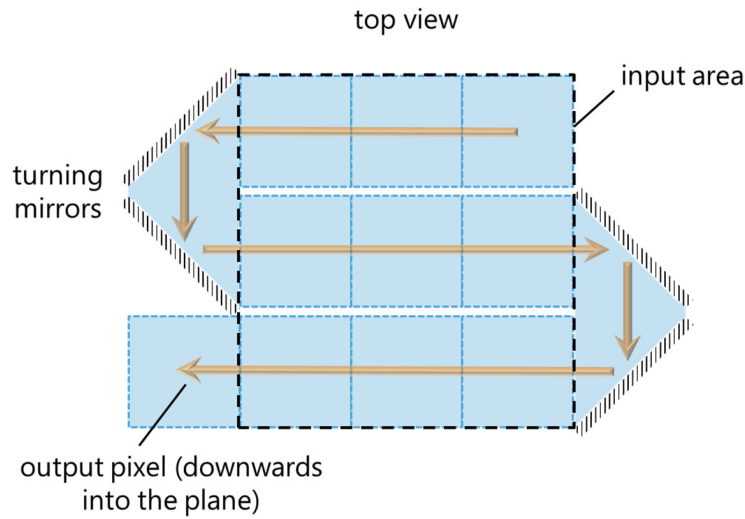

**Fig. S11. A "folded" 2-D array.** A one-dimensional line of beam splitters, as in Figs. S9 and S10, folded into a two-dimensional array, here shown in a top view. Turning mirrors chain the one-dimensional line, as in Fig. S10 A, into this two-dimensional array.

The specific illustrations here correspond to space-variant optics, with the dividing surface or transverse aperture at the far "left" of the structure. All the input light is on the right of this dividing surface or transverse aperture, and all the output light is on the left of the dividing surface or transverse aperture. So, all the light in the system flows through the aperture from right to left. Hence, we only need to consider the number, $C_{RL}$, in considering the total ONL $C$.

Hence, with this structure we have beam able to route two overlapping orthogonal input beams to the two separate output pixels. Note that in doing so we are using two horizontal "channels". The overlapping non-locality here is now $C = 2$, and two channels pass through the dividing surface or "transverse aperture" from right to left.

To repeat, it is the overlapping nature of the non-locality in this problem that requires the two separate channels, not the "degree of non-locality" – the number of beam splitters in the top row – that sets the required number of independent channels through the dividing surfaces or transverse aperture.

So far, we have shown 1-D collections of beam splitters, but it is straightforward to extend to two-dimensional arrays, at least in principle. Fig S11 illustrates a 2-D array, constructed by "snaking" a long 1-D array into a 2-D pattern, with the addition of "turning" mirrors at the ends of the rows. So, a 2-D pattern incident on the top of this array, within the approximation of dividing the pattern into such "pixels", could be collected into one output pixel. Such 2-D arrays could in principle be stacked for separating multiple orthogonal 2-D patterns.

## S12. Singular value decomposition to calculate available channels

In the main text, we have used heuristic ideas to give simple estimates of the number of available channels through the transverse aperture. Though these estimates may be quite reasonable, especially for structures much thicker than a wavelength, for relatively thin structures, they will be less accurate. Also, we have made a major simplifying assumption that, in a complex structure in which the refractive index may be varying substantially inside the region between the input and output surfaces, we can estimate the maximum number of channels based on pretending this region is filled with a uniform layer made from a material with the maximum available refractive index. Though that may seem a reasonable first estimate, it certainly can be criticized as being both approximate and, in a fundamental sense, quite unproven. A deeper use of singular value decomposition (SVD) can help with these problems, and even provide rigorous results. Here, we sketch approaches with SVD for dealing with these issues.

### Number of channels in structures with small thicknesses

With thick transverse apertures, it is reasonable, in the spirit of Fourier optics, to count the effective number plane wave functions that fit within the aperture. Though plane waves are not strictly an orthogonal basis when the aperture has finite dimensions, Fourier optics treats them as being good enough, and that is justifiable with apertures very much larger than a wavelength. With small aperture sizes (such as a small "thickness" $d$ in our notation), this approach becomes much more dubious, even if the medium between the input and output surfaces is presumed uniform.

The rigorous approach to deciding what basis functions to use is SVD analysis, which is described in detail in (*1*). Again, we could start by pretending the medium is uniform between the input and output surfaces; as a result, we know the Green's function inside the structure – it is just a "free-space" Green's function, scaled appropriately for the presumed refractive index. We could then choose some "trial" transverse aperture, so with a "trial" thickness. Then we can perform the SVD of the Green's function operator (a) between the input surface on the "right" and the transverse aperture and (b) between the transverse aperture and the output surface on the "left". Looking at the result from that, and retaining only the singular values above some minimum threshold value, we know a maximum possible $C_{RL}$ we could support. We could repeat this analysis for the "left" input to "right" output problem, similarly estimating a maximum $C_{LR}$ that could be supported.

This approach intrinsically includes all diffraction effects of the finite aperture. We could then change our "trial" aperture as needed – either to increase the number of available channels or to reduce the thickness if possible – and repeat the calculation until we came to what we regard as a reasonable compromise between too much thickness and too few channels.

## Full Green's function test of a trial design

To deal with the issue that the medium between the input and output surfaces is not uniform, we can at least check a trial design we have made, a design that has presumably led to non-uniform refractive index in the medium between the input and output surfaces. In this case, we can construct the Green's function between the input sampling points and the transverse aperture by explicitly simulating our trial design. This can construct matrices for the right-left and left-right problems, and we can then perform the SVD on these to see if the resulting matrices are supporting enough channels that are strong enough. At least this can diagnose whether we have used possibly too little thickness, giving insufficient channels, or more thickness than we need.

## Use of the eigenfunctions from SVD

When we perform SVD on a matrix, not only do we get a set of singular values, we also get corresponding sets of functions. Quite generally (*1*), SVD gives us sets of functions in the "source" or input space, i.e., $\psi_j(x,y)$, and in the "receiving" or output space, i.e., $\phi_j(u,v)$. These sets are orthogonal in their corresponding spaces, and they exist also as pairs with an associated singular value, $s_j$.

In practice, once we have chosen appropriate "sampling points" at which to represent the functions for our desired optical behavior between input and output spaces, we obtain matrices such as $\mathsf{D}_{RL}$ and $\mathsf{D}_{LR}$. The SVD of $\mathsf{D}_{RL}$ corresponds to solving the eigenproblems

$$\mathsf{D}_{RL}^\dagger \mathsf{D}_{RL} |\psi_{Rj}\rangle = |s_{RLj}|^2 |\psi_{Rj}\rangle \text{ and } \mathsf{D}_{RL} \mathsf{D}_{RL}^\dagger |\phi_{jL}\rangle = |s_{RLj}|^2 |\phi_{jL}\rangle$$

which gives the eigenvectors $|\psi_{Rj}\rangle$ and $|\phi_{Lj}\rangle$. $s_{RLj}$ are the singular values for this "right-left" problem. These eigenvectors are simply sets of function amplitudes at the sampling points in the corresponding spaces – on the "right" for the input function $|\psi_{Rj}\rangle$, and on the "left" for output function $|\phi_{Lj}\rangle$. We also have

$$\mathsf{D}_{RL} |\psi_{Rj}\rangle = s_{RLj} |\phi_{Lj}\rangle$$

We obtain similar results for $\mathsf{D}_{LR}$ with functions $|\psi_{Lj}\rangle$ and $|\phi_{Rj}\rangle$ with "left-right" singular values $s_{LRj}$. In each case, once we perform the SVD, these sets of functions are telling us specifically what orthogonal input functions we must be able to couple to what orthogonal output functions with what coupling strengths "through" the transverse aperture.

When we are trying to design the actual structure to implement some device operator D, which represents the complete desired behavior of the device, first, the SVD of D may give the most economical sets of functions to use, and may be preferable to a "dumb" choice of just the sampling points (i.e., delta functions) as the basis sets; certainly it will lead to fewer required functions in each space.

When we are trying to understand the choice of the dimensions of the transverse aperture for our device design, the SVD's of $D_{RL}$ and $D_{LR}$ are telling us explicitly just what functions we have to be able to couple "through" the transverse aperture. We can explicitly check, for a given set of transverse aperture dimensions, just how well these specific couplings could be implemented. Now looking at the problem physically with some appropriate guess or estimate for the Green's functions for the physical problem, could some input function such as some $|\psi_{Rj}\rangle$ actually couple well enough into the transverse aperture, for example, and similarly could the corresponding output function $|\phi_{Lj}\rangle$ be coupled well enough from the transverse aperture? At the very least, such an approach could allow us to diagnose why a design was not performing as well as we would like, suggesting a possible direction to improve it.

Note we are definitely saying we are only calculating the minimum number of channels required. Whether the singular value functions required map well onto the basis of functions we can physically provide is an open question. If they do not, then we may need a larger physical basis set to represent them, which would mean more channels in practice.

## S13 Comparison with a specific space-invariant kernel design

Ref. (*8*) gives specific designs for an "x times Gaussian" kernel based on a 2-D photonic crystal approach with a multilayer structure. Because it is based on a photonic crystal, then it does not support dimensional interleaving (DI) (see supplementary text S3), so we should use Eq. (3) of the main text to estimate required minimum thicknesses. Though that design is for a 2-D kernel, just as for the discussion of the imager, especially since this kernel is separable in *x* and *y*, we can usefully compare just to one "line" in the *x* direction. We can then check whether the designed structure is thick enough to support the necessary value of *C*.

As discussed in the main text, this 1-D "x times Gaussian" kernel needs $C \sim 6$. In the structure as in (*8*), $n_{max} \simeq \sqrt{2.3} = 1.516$. If we assume no restrictions on effective internal angles $\theta$, or, equivalently, on the usable fraction $\alpha$ of $k_z$ range, Eq. (3) of the main text predicts a thickness $d > 6/2n_{max} \simeq 1.98$ (free-space) wavelengths. The designed region in the middle of the preferred structure for this kernel in (*8*) is $\sim 6$ wavelengths thick, so greater than this minimum size. Overall, this is encouraging for the design of convolutional metasurfaces with moderately complex kernels (so, e.g., $C \sim 6$).

## S14. Summary of the general steps for calculating thickness limits

We can summarize the general approach to minimum thickness limits in optics and wave systems generally, all based on the number of channels *C* that must flow though the transverse aperture of the device if it is to do what we want.

1) Choose the desired function of the optics, effectively establishing the kernel or device operator *D* relating output amplitudes to input amplitudes.

2) Decide whether the design and/or fabrication approach will practically limit the usable fraction $\alpha$ or $\alpha^2$ of the corresponding 1_D or 2-D k-space or angular range inside the structure.

3) For simple, unitary (loss-less) optics, if possible, deduce the ONL *C* from dimensionality arguments and the conservation of channels, or use the SVD approach as in (4) and (5) below.

4) More generally, construct the matrix representation D of *D*, using either (a) pixels for functions that are already discretized or (b) a sufficiently dense sampling of any continuous functions, noting that numerical aperture restrictions define a useful separation $\delta l$ (Eq. (11) of the main text) of sampling points.

5) Choose the position of a dividing surface, construct the truncated "right-left" and (if necessary) "left-right" matrices $D_{RL}$ and $D_{LR}$, and perform their SVDs, noting the total number *C* of singular values of both SVDs that lie above some practical minimum magnitude. If necessary, move the dividing surface to find the largest value of the ONL *C*.

6) Decide whether the design and/or fabrication approach supports dimensional interleaving. If it does, use Eq. (4) of the main text to estimate the minimum area of the transverse aperture, and deduce the minimum thickness *d* given some width or diameter *L* of the optics. If not, use Eq. (3) of the main text to estimate the minimum thickness *d*.

## S15. Additional discussion

A few additional points are worth clarifying and emphasizing here.

### Nonlocality itself does not require thickness

We need to be clear that it is not nonlocality itself that necessitates large numbers of "sideways" channels. Arbitrary nonlocality is possible in a "single mode" device. For example, consider a single-mode waveguide with a number of "taps" that couple light in or out. There is no particular limit to how many taps we put on one waveguide. Shining light into the waveguide input would give light out of every tap and any remaining light out of the far end. If we run this structure "backwards", with the phase conjugate of all the emitted light injected back into the taps and the far end, then that light will be coherently combined to come back out the waveguide "input". Hence, we see that arbitrary nonlocality (multiple input/output taps) is possible for just one "mode" that crosses a dividing surface at the fiber input end. It is the *overlapping* nature of nonlocality for multiple such outputs that necessitates multiple modes across the dividing surface, not just nonlocality itself. (We have also given a simple example of the difference between non-locality and overlapping non-locality in supplementary text S11.)

### Two-dimensional kernels

We illustrated the SVD approach here explicitly mostly for just 1-D kernels, but we can extend to 2-D kernels (and we gave one simple example approach in supplementary text S11). We would keep the dividing surface still in one direction (e.g., *y*), but extend the kernel to be 2-D, leading to a correspondingly larger matrix for which to find the singular values, but the approach would otherwise be the same.

## Complex kernels

For simplicity of illustration, in this paper we show results for real kernels. However, no changes are required to the mathematics if the kernels are complex, and such kernels are routine in optics to represent phases other than 0 or 180°. Then we will be looking at the magnitude of any singular values to decide if they are above some minimum threshold when we are counting the number of channels we need.